\documentclass[sigconf]{acmart}

\usepackage{graphicx}

\IfFileExists{headers/config/showoverfull.config}{
	\overfullrule=1cm
}{
}

\usepackage{etoolbox}

\newbool{includeappendix}
\setbool{includeappendix}{true} %
\IfFileExists{headers/config/noappendix.config}{
	\setbool{includeappendix}{false}
}{}

\usepackage{xr} %
\usepackage{filecontents}

\ifbool{includeappendix}{}{
	\input{appendix-labels-loader}

	\externaldocument{appendix-labels}
}

\newcommand{\eg}{e.g., }
\newcommand{\ie}{i.e., }

\usepackage{acro} %

\DeclareAcronym{cli} {
    short = CLI,
    long = Command Line Interface,
}

\usepackage{subcaption}

\usepackage{xcolor} %

\definecolor{lightblue}{RGB}{212,234,248}
\definecolor{darkblue}{RGB}{29,117,180}
\definecolor{lightgreen}{RGB}{204,255,204}
\definecolor{darkgreen}{RGB}{101,178,50}
\definecolor{orange}{RGB}{255,171,89}

\definecolor{w1color}{rgb}{0.2980392156862745, 0.4470588235294118, 0.6901960784313725}
\definecolor{w2color}{rgb}{0.8666666666666667, 0.5176470588235295, 0.3215686274509804}
\definecolor{w3color}{rgb}{0.3333333333333333, 0.6588235294117647, 0.40784313725490196}

\usepackage{listings}

\usepackage{textcomp}

\usepackage{xcolor}

\usepackage[scaled=0.8]{beramono}

\definecolor{ckeyword}{HTML}{7F0055}
\definecolor{ccomment}{HTML}{3F7F5F}
\definecolor{cstring}{HTML}{2A0099}

\lstdefinestyle{numbers}{
	numbers=left,
	framexleftmargin=20pt,
	numberstyle=\tiny,
	firstnumber=auto,
	numbersep=1em,
	xleftmargin=2em
}

\lstdefinestyle{layout}{
	frame=none,
	captionpos=b,
}

\lstdefinestyle{comment-style}{
	morecomment=[l]//,
	morecomment=[s]{/*}{*/},
	commentstyle={\color{ccomment}\itshape},
}

\lstdefinestyle{string-style}{
	morestring=[b]",%
	morestring=[b]',%
	stringstyle={\color{cstring}},
	showstringspaces=false,%
}

\lstdefinestyle{keyword-style}{
	keywordstyle={\ttfamily\bfseries},
	morekeywords={
		function,
		constructor,
		int,
		bool,
		return,
		returns,
		uint
	},
	morekeywords = [2]{},
	keywordstyle = [2]{\text},
	sensitive=true,
}

\lstdefinestyle{input-encoding}{
	inputencoding=utf8,
	extendedchars=true,
	literate=
	{ℝ}{$\reals$}1%
	{→}{$\rightarrow$}1%
	{α}{$\alpha$}1%
	{β}{$\beta$}1%
	{λ}{$\lambda$}1%
	{θ}{$\theta$}1%
	{ϕ}{$\phi$}1%
}

\lstdefinestyle{escaping}{
	moredelim={**[is][\color{blue}]{\%}{\%}},
	escapechar=|,
	mathescape=true
}

\lstdefinestyle{default-style}{
	basicstyle=\fontencoding{T1}\ttfamily\footnotesize,
	style=numbers,
	style=layout,
	style=comment-style,
	style=string-style,
	style=keyword-style,
	style=input-encoding,
	style=escaping,
	tabsize=2,
	upquote=true
}

\lstdefinelanguage{BASIC}{
	language=C++,
	style=default-style
}[keywords,comments,strings]%

\usepackage{tikz}

\usetikzlibrary{arrows}
\usetikzlibrary{automata}
\usetikzlibrary{calc}
\usetikzlibrary{backgrounds}
\usetikzlibrary{decorations.markings}
\usetikzlibrary{decorations.pathmorphing}
\usetikzlibrary{decorations.pathreplacing}
\usetikzlibrary{fit}
\usetikzlibrary{patterns}
\usetikzlibrary{positioning}
\usetikzlibrary{shadows}
\usetikzlibrary{shapes}
\usetikzlibrary{shapes.geometric}

\microtypecontext{spacing=nonfrench}

\usepackage[ruled,vlined,linesnumbered]{algorithm2e}
\usepackage{setspace}

\usepackage[normalem]{ulem} %
\usepackage{amsmath,amsthm}
\usepackage{booktabs}
\usepackage{xspace}
\usepackage{enumitem}
\usetikzlibrary{backgrounds}
\usepackage{stackengine}
\usepackage[shortcuts]{extdash}  %
\usepackage[most]{tcolorbox}

\usepackage{siunitx}

\usepackage{placeins}  %
\usepackage{layouts}  %

\usepackage{balance}  %
\usepackage{fancyhdr}  %

\definecolor{orange}{RGB}{255,153,51}
\definecolor{lightorange}{RGB}{255,235,214}
\definecolor{darkorange}{RGB}{171,68,1}

\definecolor{lightgrey}{RGB}{242,242,242}
\definecolor{midgrey}{RGB}{191,191,191}

\definecolor{darkblue}{RGB}{29,117,180}
\definecolor{lightblue}{RGB}{205,224,220}

\definecolor{darkred}{RGB}{255,32,33}
\definecolor{lightred}{RGB}{255,128,129}

\definecolor{darkgreen}{RGB}{101,178,50}
\definecolor{lightgreen}{RGB}{170,223,135}

\definecolor{pastelblue}{rgb}{0.6588235294117647, 0.8392156862745098, 1}

\usepackage{mdframed}

\global\mdfdefinestyle{lightgrey}{%
    backgroundcolor=lightgrey,
    linewidth=0pt,
    shadow=true,
    shadowcolor=midgrey,
    shadowsize=1pt,
    skipabove=0.5em,
    skipbelow=1em,
    innertopmargin=0.5em,
    innerbottommargin=0.5em,
    innerleftmargin=0.5em,
    innerrightmargin=0.5em,
}

\DeclareMathOperator*{\softmax}{soft\,max}

\newtheorem{definition}{Definition} %

\newcommand{\name}{\textsf{Memento}\xspace}

\DeclareSIUnit{\nothing}{\relax}
\newcommand*{\kilonum}[1]{\SI{#1}{\kilo\nothing}}

\newcommand*{\meganum}[1]{\SI{#1}{\mega\nothing}}

\SetKwInput{Parameter}{Parameters}

\newcommand{\first}{\emph{(i)}\xspace}
\newcommand{\second}{\emph{(ii)}\xspace}
\newcommand{\third}{\emph{(iii)}\xspace}

\newcommand{\capt}[1]{\textsl{\small #1}}

\newcommand{\fugu}{\textsf{Fugu}\xspace}
\newcommand{\fugufeb}{\textsf{Fugu}\textsubscript{Feb}\xspace}

\makeatletter
\renewcommand\paragraph{\@startsection{paragraph}{4}{\z@}%
    {1.50mm}%
    {-1em}%
    {\normalfont\normalsize\bfseries}}
\makeatother

\newcommand{\remove}[1]{}

\usepackage[capitalize]{cleveref}

\makeatletter
\newcommand\theHALG@line{\thealgorithm.\arabic{ALG@line}}
\makeatother

\crefformat{section}{\S#2#1#3}

\crefrangeformat{section}{\S#3#1#4\crefrangeconjunction\S#5#2#6}

\crefmultiformat{section}{\S#2#1#3}{\crefpairconjunction\S#2#1#3}{\crefmiddleconjunction\S#2#1#3}{\creflastconjunction\S#2#1#3}

\newcommand{\crefrangeconjunction}{--}

\crefname{listing}{Lst.}{listings}
\crefname{line}{Lin.}{Lin.}
\crefname{appendix}{Appendix}{Appendix}

\newcommand{\appref}[1]{%
	\ifbool{includeappendix}{\cref{#1}}{the appendix}%
}
\newcommand{\Appref}[1]{%
	\ifbool{includeappendix}{\cref{#1}}{The appendix}%
}

\setcopyright{none}
\renewcommand\footnotetextcopyrightpermission[1]{}

\fancypagestyle{plain}{%
    \fancyhf{} %
}
\fancypagestyle{firstpagestyle}{%
    \fancyhf{} %
}

\begin{document}

\title{
    On Sample Selection for Continual Learning:\\
    a Video Streaming Case Study
}
\date{}

\author{Alexander Dietmüller}
\affiliation{\institution{ETH Zürich, Switzerland}}
\email{adietmue@ethz.ch}
\orcid{0000-0003-3769-3958}

\author{Romain Jacob}
\affiliation{\institution{ETH Zürich, Switzerland}}
\email{jacobr@ethz.ch}
\orcid{0000-0002-2218-5750}

\author{Laurent Vanbever}
\affiliation{\institution{ETH Zürich, Switzerland}}
\email{lvanbever@ethz.ch}
\orcid{0000-0001-7419-2971}

\begin{abstract}
    Machine learning (ML) is a powerful tool to model the complexity of communication networks.
    As networks evolve, we cannot only train once and deploy.
    Retraining models, known as continual learning, is necessary.
    Yet, to date, there is no established methodology to answer the key questions:\\
    \emph{With which samples to retrain?} \hfill \emph{When should we retrain?}

    We address these questions with the sample selection system \name, which maintains a training set with the “most useful” samples to maximize sample space coverage.
    \name particularly benefits rare patterns---the notoriously long ``tail'' in networking---and allows assessing rationally \emph{when} retraining may help, \ie when the coverage changes.

    We deployed \name on Puffer, the live-TV streaming project, and achieved a \SI{14}{\percent} reduction of stall time, $3.5\times$ the improvement of random sample selection.
    Since \name does not depend on a specific model architecture, it is likely to yield benefits in other ML-based networking applications.
\end{abstract}



\maketitle

\section{Introduction}
\label{sec:introduction}

Adaptive Bit Rate (ABR) algorithms aim to avoid video stalls while maximizing the image quality in video streaming.
This entails predicting the transfer time of video chunks, a complex task for which researchers are increasingly using machine learning (ML)~\cite{yanLearningSituRandomized2020,maoNeuralAdaptiveVideo2017,xiaAutomaticCurriculumGeneration2022}.
Current ML-based ABR algorithms already achieve good average performance; \ie high image quality while largely avoiding stalls~\cite{yanLearningSituRandomized2020, langleyQUICTransportProtocol2017}.
But rare stalls matter as they are known to impact user experience far more than image quality~\cite{duanmuQualityofExperienceIndexStreaming2017,krishnanVideoStreamQuality2013}.
Moreover, as networks evolve, maintaining performance over time is a concern that hinders the deployment of ML-based solutions.
This challenge is known as continual learning~\cite{parisiContinualLifelongLearning2019,buzzegaRethinkingExperienceReplay2020,iseleSelectiveExperienceReplay2018,rolnickExperienceReplayContinual2019}.

In this paper, we ask if and how we can improve tail performance over time while maintaining the average. For ABR, this translates to reducing stalls while maintaining quality.

\begin{figure}[t]
  \centering
  \includegraphics[scale=0.48]{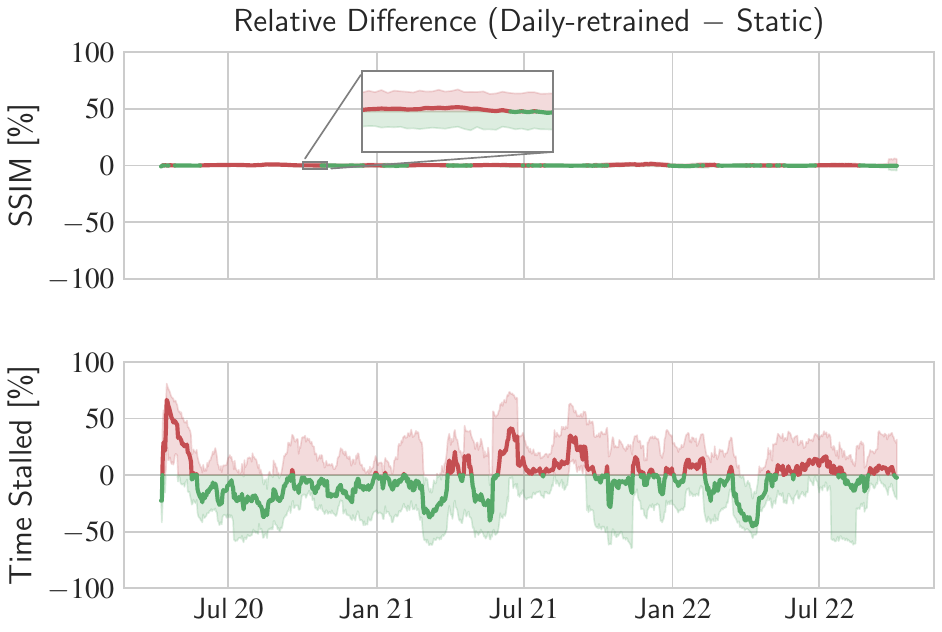}
  \caption{%
    On Puffer~\cite{Puffer}, retraining daily with random samples did not outperform a never-retrained model consistently.
    On average, image quality and stream-time spent stalled differ by less than \SI{0.2}{\percent} and \SI{4.2}{\percent}.\\
    \capt{\footnotesize Mean and 90\% CI over a one-month sliding window.
      Data source:~\cite{Puffer}}
  }
  \label{fig:puffer_motivation}
\end{figure}

The most common approach to improve ML performance is \textbf{using more}: more data, more training, more complex models.
However, this does not guarantee (tail) improvements.
\textbf{More training} does not help if done using the wrong samples.
Similarly, naively using \textbf{more data} does not improve tail performance if it does not address the imbalance between average and tail samples.
Finally, \textbf{more models} can help to address this if they cover different parts of the data distribution.
Otherwise, the individual models fail to complement each other and provide no benefit.
Besides its uncertain effectiveness, \textbf{using more} increases training and inference time.
For large networking applications, this can be significant.
If YouTube were to use ML-based ABR, it would require inference for $\approx$30 billion video chunks per day~\cite{YouTubeStatistics20242024}:
slower inference means higher costs in delay, energy, and money.

Instead of \textbf{using more}, we propose to improve performance with a \emph{smarter selection} of training samples,
which we motivate with the following case study.

\paragraph*{Puffer case study}
The best study of continual learning in networking to date is Puffer~\cite{Puffer}. This ongoing study monitors ABR performance with users streaming live TV with randomly assigned algorithms.
Puffer's authors proposed their own ML-based ABR and retrained it daily using random samples from the past two weeks~(\cref{fig:puffer_motivation}).

To the author's surprise~\cite{yanLearningSituRandomized2020}, retraining every day brought essentially no benefits: Over almost \num{900} days, it improved image quality by only \SI{0.17}{\percent} over a static---never retrained---version~(\cref{fig:puffer_motivation}, top).
On the tail, daily retraining reduced the fraction of stream time spent stalled by \SI{4.17}{\percent} on average, with large fluctuations over time~(\cref{fig:puffer_motivation}, bottom).
This illustrates the \textbf{more training} approach falling short.
But \emph{why?}
Why is the daily-retrained model not consistently outperforming the static one?
Why does it stall only half the time in some months but twice as much in others~(\cref{fig:puffer_motivation}, bottom)?

\begin{figure}[t]
  \centering
  \includegraphics[scale=0.5]{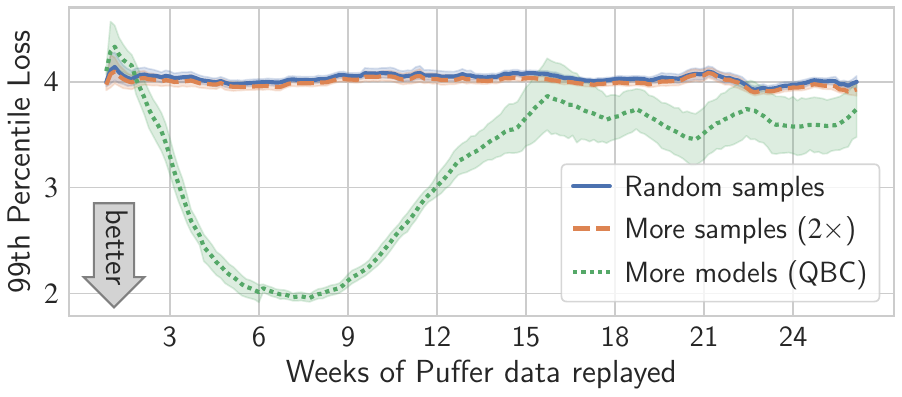}
  \caption{
    More resources do not guarantee better tail performance.
    Using more random samples has virtually no effect.
    Using more models to select training samples (QBC) initially helps but degrades over time.\\
    \capt{\footnotesize Mean and 90\% CI over a two-week sliding window
      (see \cref{sec:puffer} for details).}
  }
  \label{fig:overview_more}
\end{figure}

We argue retraining did not help because training samples were selected randomly.
As most streaming sessions perform similarly, this leads to an imbalanced training set with many similar samples and few tail ones.
Doubling the number of training samples in the same setup---\textbf{more data}---does not help either~(\cref{fig:overview_more}):
we must address this imbalance to improve a model's tail performance.
Query-By-Committee (QBC)~\cite{seungQueryCommittee1992} is a classic approach to address such imbalance by selecting samples where a committee of models disagrees and prediction entropy is high.
When applying QBC to the Puffer data, we observe that the models fail to identify rare samples reliably and eventually overfit on noise~(\cref{fig:overview_more}), \ie using \textbf{more models} is not enough to improve tail performance.

\paragraph*{Problem}
\cref{fig:puffer_motivation} and \cref{fig:overview_more} suggest that retraining \emph{can} improve tail performance, but we observe common approaches like random sampling or QBC to be ineffective or unreliable over time.
If the selection got lucky, retraining improved the tail; other times, retraining was useless or even detrimental.
We argue that we can do better with a smarter selection strategy that reliably picks up important tail samples.
In summary, we must answer the following two questions:

\begin{enumerate}[nosep, topsep=0.5mm,itemsep=1mm]

  \item From a stream of new samples, \emph{with which samples should we retrain} to improve performance over time?

  \item From an updated set of training samples in memory, \emph{when should we retrain the model?}
        Given the large resource costs, can we avoid retraining ``for nothing''?

\end{enumerate}

\noindent
Fundamentally, an algorithm addressing these questions requires:
\first a signal to \emph{select important samples}, able to identify the tail;
\second a signal to \emph{quantify changes} to decide when to retrain;
\third a mechanism to \emph{forget} noise and outdated samples to avoid degradation over time (QBC in \cref{fig:overview_more}).
Finally, resource usage should be small compared to \textbf{using more}.

\begin{figure}[b]
  \centering
  \includegraphics[scale=0.88]{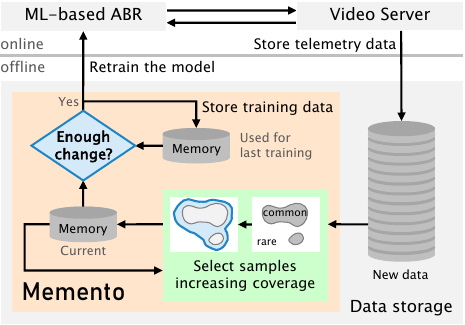}
  \vspace*{-3mm}
  \caption{
    \name maximizes sample space coverage,
    improving the tail while rationalizing when to retrain.
  }
  \label{fig:overview_retraining}
\end{figure}

\paragraph*{Solution}
\hspace{-4.7pt}We propose a sample-space-aware continual learning algorithm based on \emph {coverage maximization}, \ie prioritizing samples from low-density areas of the sample space, and present \name, a prototype implementation of this idea.
\cref{fig:overview_retraining} illustrates how \name integrates with an ML-based video streaming system transmitting videos in chunks.
In ML-based ABR, a \emph{sample} encompasses telemetry data for recent video chunks; samples are collected by the server.

\name estimates the sample-space density from both new and in-memory samples, prioritizing rare low-density samples to address dataset imbalance and improve tail performance.
\name uses the \emph{difference in coverage} to assess whether
there is something new to learn. If so, it retrains.

\paragraph{Main contributions}
\begin{itemize}[nosep, topsep=0.5mm,itemsep=0.5mm]
  \item[\textbf{\cref{sec:overview}}]
    We propose a sample-space-aware continual learning algorithm using density to improve tail performance.
  \item[\textbf{\cref{sec:memory}}]
    We present \name~(\cref{fig:overview_retraining}), a prototype implementation that uses sample-space density for sample selection.
    \name is publicly available~(\Cref{app:artefacts}).
  \item[\textbf{\cref{sec:puffer}}]
    We validate the effectiveness of our algorithm by applying \name on an extensive ABR case study.
    \begin{itemize}[nosep]
      \item We deployed \name on Puffer and collected \num{10} stream-years of real-world data over \num{9} months.
      \item \name stalls \SI{14}{\percent} less than the static model (\num{3.5}$\times$ better than daily retraining with random samples).
      \item \name achieves this by retraining only \num{7} times with a marginal degradation on image quality~(\SI{0.13}{\percent}).
      \item \name has easy-to-tune parameters.
    \end{itemize}
  \item[\textbf{\cref{sec:simulation}}]
    We conduct microbenchmarks using data center workloads to show that out algorithm is not limited to ABR.
\end{itemize}

\section{A case for density}
\label{sec:overview}

We propose a continual learning algorithm based on estimating sample space density.
Estimating density allows the algorithm to prioritize rare low-density samples, ultimately maximizing the sample space coverage.
This section provides intuition as to why we propose this selection metric instead of per-sample metrics like sample loss or QBC's entropy.

\paragraph{Density for sample selection}
``The tail'' is not a single pattern that rare traffic follows, but rather many patterns with only a few samples each.
A random sample selection mirrors any imbalance in the underlying distribution, \eg over-represented common traffic patterns.
This leads to \emph{diminishing returns} as we sample more and more from common patterns and little from the tail.
As illustrated by \cref{fig:puffer_motivation}, this yields good average performance but is unreliable at the tail.
We need to address the imbalance of the training set.

To correct dataset imbalance, we need a sample-space-aware selection.
Traditional approaches use the model performance (\eg entropy, loss, reward) to select samples~\cite{buzzegaRethinkingExperienceReplay2020,iseleSelectiveExperienceReplay2018}.
We found that this does not work well on Puffer because it fails to \emph{avoid catastrophic forgetting}.
By considering each sample independently, they fail to preserve samples with good performance representing `normal' traffic.

Instead, we propose to select samples based on the density of their neighborhood, which considers the whole sample space:
a \emph{sample-space-aware selection} that aims to \emph{maximize coverage} of the sample space.
The key insight to maximize coverage is to retain samples where few similar samples exist.
We achieve this by removing samples from high-density areas with the most similar samples.
This decreases the density and we naturally stop removing further samples.

\begin{figure}[t]
  \centering
  \includegraphics[scale=0.5]{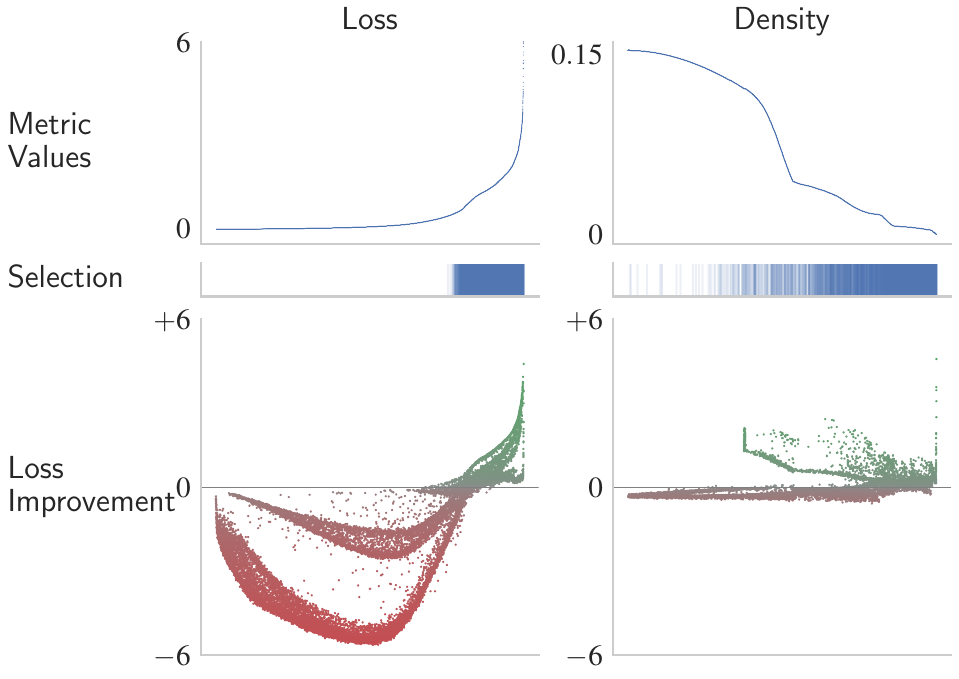}
  \caption{%
    \capt{
      Loss improvement obtained by retaining with 1M samples over a dataset of 5M.
      The same batches of 256 samples are used for the loss- (left) and density-based selection (right).
    }\\
    To improve tail performance, we need many low-density batches because
    they are all different.
    To maintain average performance, we need only a few high-density batches, as they are similar.
    Selecting based on density (right) achieves both.
    Conversely, loss-based selection (left) is too specific.
    It suffers from diminishing returns by selecting too many high-loss batches and catastrophically forgets the average.
  }
  \label{fig:loss_selection_matrix}
\end{figure}

\cref{fig:loss_selection_matrix} illustrates the benefits of samples-space-aware density versus sample-aware loss as a selection metric.
In this experiment, we use the static model from Puffer's authors and \meganum{5} samples collected by Puffer over a few days.
We create batches of 256 samples and compute their loss and density (as detailed in~\cref{sec:memory}).
The top row of \cref{fig:loss_selection_matrix} shows the mean loss and density per batch.
We then use each metric to select \meganum{1} samples and retrain the model.
The bottom row compares the effect of each selection metric over the \meganum{5} samples by showing the mean loss improvement per batch.

The left column shows that loss-based selection focuses \emph{too much} on the tail, \ie high-loss batches.
It yields good improvements there but does not preserve enough average samples and degrades performance on most other batches.

Conversely, the right column illustrates that density-based selection is more holistic, focusing on the tail while covering the entire sample space.
Performance is improved at the tail, \ie on low-density batches, without much degradation on the high-density ones.
One may object that the loss selection could be tuned to maintain more ``low-loss samples.'' We tried this and our evaluation shows that it does not perform as well as density and can easily get much worse~(\cref{ssec:benchmarks,fig:alt_loss}).

These results also shed some light on why we observed QBC degrade over time (\cref{sec:introduction}, \cref{fig:overview_more}).
Overall, QBC has some inertia, as models are gradually replaced (more details in \cref{ssec:benchmarks}), but we ultimately observe a similar effect as loss-based selection.
Noisy samples are, by definition, random and have high entropy.
Consequently, they are preferred by QBC.
Over time, QBC accumulates noise and forgets the average and more common tail samples.
As a result, it is unable to maintain the initial performance improvements on the tail.

\paragraph{Density for shift detection}
Continual learning aims to adapt when the data distribution changes.
These changes can be broadly categorized into \emph{covariate shift}~\cite{shimodairaImprovingPredictiveInference2000}, \ie previously unseen traffic patterns emerge or the prevalence of patterns changes and \emph{concept drift}~\cite{widmerLearningPresenceConcept1996}, \ie the underlying network dynamics change.
Density-based sample selection is an effective tool to capture both kinds of changes.

When new traffic patterns appear, \eg users starting to stream over satellite networks, they populate a previously-empty area of the sample space, resulting in a low density and, thus, a high probability of being selected for retraining.
Similarly, patterns becoming less prevalent results in low density, and a high probability of remaining selected.

When the underlying network dynamics change, \eg a new congestion control gets deployed, we should forget old samples that are no longer relevant. However, detecting those changes is difficult, and being wrong risks forgetting useful information such as hard-to-gather tail samples.

Using density for sample selection \emph{correlates the probability of forgetting samples with how important it is to remember them.}
Samples from dense regions are discarded readily, as we will likely get more of those samples.
Conversely, low-density batches are less likely discarded as we only encounter these samples infrequently.
This makes the selection more conservative at the tail, allowing to remember tail patterns; the model will perform well on similar traffic if it recurs. If it does not, \ie it was essentially noise, then it will eventually be forgotten.
We provide some empirical evidence of recurring patterns in the tail of the Puffer traffic in \cref{app:puffer_comparison_extra}.

\vfill
\begin{algorithm}[b]
  \small
  \setstretch{1.1}
  \caption{\name Coverage Maximization}
  \label{alg:coverage}
  \newcommand*{\whitespace}{%
    \vspace*{4mm}
}

\Parameter{%
    Capacity $C$, %
    threshold $\tau$,\linebreak
    batch size $b$, %
    bandwidth $h$, %
    temperature $T$ %
}
\KwIn{%
    In-memory $S_{mem}$ and incoming $S_{new}$ samples
}
\KwOut{%
    Selected samples $S^*$, decision $retrain$
}
\Begin{%
    $S^* \leftarrow S_{new} \cup S_{mem}$\\
    $B' \leftarrow \{\}$
    \tcp*{or last train batches}
    $\textit{retrain} \leftarrow \textit{False}$\\

    \whitespace
    \tcp{\cref{ssec:distances}:\ Distance measurement}
    $B \leftarrow \textit{BatchSamples}(S^*, b)$\\
    $B \leftarrow BBDR(B)$
    \tcp*{$(x,y) \to (\hat{y},y)$}
    $(D^{pred}, D^{out}) \leftarrow \textit{DistributionDistances}(B)$\\

    \whitespace
    \tcp{\cref{ssec:density}:\ Density estimation}
    $\widehat{\rho}^{\,k} \leftarrow KDE(D^{k},h)
        \hfill \forall k\in{\{pred,out\}}$\\
    $\widehat{\rho} \leftarrow \min (\widehat{\rho}^{\,pred}, \widehat{\rho}^{\,out})$\\

    \whitespace
    \tcp{\cref{ssec:selection}:\ Sample selection}
    \While{$|S^*| > C$}{\label{line:while}
        $p^{discard} \leftarrow \softmax(\widehat{\rho}\,/\,T)$\\
        $i \leftarrow \textit{WeightedRandomChoice}(p^{discard})$
        \label{line:max_density}\\

        $S^*, B \leftarrow \textit{DiscardBatch}(S^*, B, i)$\\
        $\widehat{\rho}, D \leftarrow\textit{UpdateDensities}(\widehat{\rho}, D, i)$
        \label{line:update}\\

    }

    \whitespace
    \tcp{\cref{ssec:changes}:\ Retraining decision}
    \If{$RCI(B, B') \geq \tau$}{
        $\textit{retrain} \leftarrow \textit{True}$\\
        $B' \leftarrow B$
        \tcp*{remember train batches}
    }
}

\end{algorithm}

\pagebreak
\section{Coverage maximization}
\label{sec:memory}

The core of \name is \Cref{alg:coverage}: its sample selection to approximately maximize the sample-space coverage.

\name achieves this by estimating the sample space density from the distances between samples:
the higher the density of a point in sample space (\ie the more samples are close to it) the better the memory covers this part of sample space.
\name approximates optimal coverage by iteratively discarding samples, assigning a higher discard probability to high-density regions.
It proceeds in four steps:

\begin{enumerate}[nosep,topsep=2mm, itemsep=2mm]

  \item It computes pairwise distances between distributions of sample batches, leveraging batching and  \emph{black-box dimensionality reduction} (BBDR) for scalability~(\cref{ssec:distances});

  \item It estimates input- and output-space density using \emph{kernel density estimation} (KDE, \cref{ssec:density}).

  \item It discards batches \emph{probabilistically} until fitting the memory constraints.
        It maps density to discard probability, balancing tail-focus and noise rejection~(\cref{ssec:selection}).

  \item Once new samples are selected, it approximates how much the memory coverage has increased to decide whether retraining might be beneficial~(\cref{ssec:changes}).
\end{enumerate}

\noindent
\name's sample selection relies on three internal parameters:
the batch size $b$, the KDE bandwidth $h$, and the probability mapping temperature $T$.
We provide default choices for these parameters in \cref{ssec:puffer_results} and analyze the impact of each parameter on \name's performance in \cref{ssec:benchmarks}.

\subsection{Definitions}
\label{ssec:problem_statement}

\paragraph{Process}
We consider a process $y = f (x)$ that maps inputs $x \in \mathcal{X} = \mathbb{R}^n$ to outputs $y \in \mathcal{Y}$, where $\mathcal{Y} = \mathbb{R}$ in regression or $\{1,2,\dots,k\}$ in classification problems.
In the context of ABR, $f$ models the network dynamics mapping traffic features $x$ (\eg video chunk size, TCP statistics, transmission times of past packets) to a prediction for the next chunk $y$ (\eg the current bandwidth or expected chunk transmission time).

\paragraph{Predictions}
We consider a model $\hat{f}$ that is trained to predict $\hat{y} = \hat{f} (x)$ from a set of training samples $S$, where \linebreak
${S = \{ (x_1, y_1), \dots, (x_N, y_N))\}}$, \ie a supervised setting.

\paragraph{Replay memory}
To account for concept drift or covariate shift, we retrain $\hat{f}$ with an updated set of samples $S^*$, stored in a \emph{replay memory} with capacity $C$.
A sample selection strategy decides how to update this memory.

\paragraph{Sample selection strategy}
Given a set $S_{new}$ of new samples available and a set $S_{mem}$ of samples currently stored in memory, with $|S_{new}| + |S_{mem}| > C$, a selection strategy must select $S^*\subset (S_{new} \cup S_{mem})$, such that $|S^*| \leq C$.

\subsection{Distance measurement}
\label{ssec:distances}

\paragraph{Batching}
Before discussing the distance computation, we need to consider scalability.
\cref{alg:coverage} has two main bottlenecks:
\first computing pairwise distances ($\mathcal{O}(n^2)$);
and \second updating density estimates after discarding a sample ($\mathcal{O}(n)$).
Puffer~\cite{Puffer} currently collects over \meganum{1} samples daily, and processing each sample individually would consume a prohibitive amount of resources for \name to be practical.

\name scales by aggregating samples in batches, computing distances between aggregates, as well as discarding samples and updating densities for an entire batch at once.

Batching improves scalability but reduces the flexibility of sample selection.
For example, if a common and a rare sample are aggregated in the same batch, they can only be kept or discarded together.
To avoid suboptimal aggregation, samples are first grouped by outputs, then by predictions, and finally split into batches; this pools samples spatially to create homogeneous aggregates.
We found this approach to work best, but \name also supports batching based on sampling time or application-specific criteria.

\begin{figure}[t]
  \centering
  \includegraphics[scale=0.5]{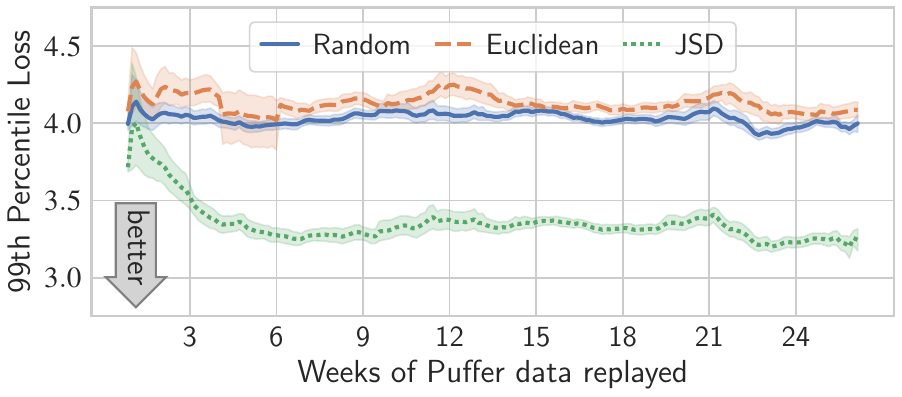}
  \caption{%
    Selecting samples based on the Euclidean distance between batch averages does not improve tail performance.
    Computing the Jensen-Shanning distance between batch distributions is a better choice.\\
    \capt{\footnotesize Mean and 90\% CI over a two-week sliding window
      (see \cref{sec:puffer} for details).}
  }
  \label{fig:distance_comparison}
\end{figure}

\paragraph{Distribution distances}
A natural first choice for batch aggregation is to average sample vectors and compute the Euclidean distance between the averages.
However, we find that this approach is flawed: averaging ``dilutes'' rare values in a batch and the resulting distance is not sensitive to tail differences. \cref{fig:distance_comparison} shows that selecting samples based on the Euclidean distance between batch averages does not improve tail performance compared to a random sample selection.

Instead, we aggregate batches by computing an empirical sample distribution\footnote{This may also be interpreted as replacing distances between observations (samples) with distances between the underlying processes (distributions).}.
This preserves rare values, but we still need to reduce the dimension to gain any computational benefit from batching.
We address this problem using \emph{black-box dimensionality reduction} (BBDR), an approach inspired by research on dataset shift detection~\cite{liptonDetectingCorrectingLabel2018} that has been found to outperform alternatives like PCA~\cite{rabanserFailingLoudlyEmpirical2019}.
The idea is simple:
Use the existing model---trained to identify important feature combinations---to predict $\hat{y}$ (the predicted transmission time), then compute distances in the low-dimensional prediction space.
In essence, BBDR leverages prediction as dimensionality reduction tailored to the task at hand.

Finally, we compute the distance between batch distributions using the Jensen-Shannon Distance (JSD)~\cite{endresNewMetricProbability2003} which is closely related to the Kullback-Leibler Divergence (KL)~\cite{kullbackInformationSufficiency1951}.
The KL is an attractive choice to improve tail performance as it puts a large weight on rare values, \ie values with high probability in one distribution but not in the other.
However, the KL has two drawbacks. It is:
\first asymmetric, only sensitive to rare values in one distribution;
and \second only a divergence, while common density estimation methods require a distance (see below).
Thus, we choose the closely related JSD, which is both symmetrical and a distance metric~\cite{endresNewMetricProbability2003}:

\begin{definition}[Jensen-Shannon Distance]
  Let $P$ and $Q$ be probability distributions, and let $M = \tfrac{1}{2} (P + Q)$.
  The Jensen-Shannon Distance between $P$ and $Q$ is defined by:
  \begin{align*}
    JSD(P, Q) & = \sqrt{\tfrac{1}{2} \left( KL(P, M) + KL(Q, M) \right)}                  \\
    \text{where} \qquad
    KL(P, M)  & = \sum_{x\in\mathcal{X}} P(x) \log_2 \left(\frac{P(x)}{M(x)}\right)       \\
    \intertext{if $P$ and $Q$ are discrete distributions in the space $\mathcal{X}$; or}
    KL(P, M)  & = \int_{-\infty}^{\infty} p(x) \log_2 \left(\frac{p(x)}{m(x)}\right)\, dx
  \end{align*}%
  if $P$ and $Q$ are continuous probability distributions with probability density functions $p$ and $q$, and $m = \tfrac{1}{2} (p + q)$.%
\end{definition}

\paragraph{Inputs and Outputs}
It is not clear a priori whether distances in the sample input or output space are more important for tail performance.
Thus, in supervised classification or regression problems---where outputs are available---\name considers both equally and computes separate distances for both spaces, which are combined below~(\cref{ssec:density}).

\paragraph{Probabilistic predictions}
During batching, \name \linebreak leverages probabilistic predictions; \eg the transition time predictor in Puffer~\cite{yanLearningSituRandomized2020} outputs a probability distribution over 21 transition time bins.
This probability distribution captures whether predictions are somewhat uncertain---thus indicating a need for training using more such samples---or certain.
However, we must handle probabilistic estimates differently than point estimates:
\first When batching samples, \name groups probabilistic predictions first by the distribution mode, then by their probability;
\second The batch distribution is computed as a mixture distribution, \ie
the prediction distribution for batch $b_i$ is $P_i(x) = ({1}/{|b_i|})\cdot\sum P_j(x)$, where $P_j$ is the probabilistic prediction for sample $j$.

\subsection{Density estimation}
\label{ssec:density}

From the sample distances, we can compute the prediction- and output-space densities.
Since we do not know the topology of the sample space, we cannot use simple approximations, such as the fraction of samples per cluster.
A common general approach is kernel density estimation~(KDE)~\cite{parzenEstimationProbabilityDensity1962}.

\begin{definition}[KDE]
  Let $b$ be a sample batch and $B$ a set of sample batches, and
  let $d^k(b,\, b')= JSD(P^k,\,{P^k}')$ with $k\in\{pred, out\} $ be the prediction or output distance between batches $b$ and $b'$ with distributions $P^k,\,{P^k}'$.
  Then, using a Gaussian kernel with \emph{bandwidth} $h$, the kernel density estimate $\hat{\rho}$ at the location of batch $b$ is defined as:
  \begin{align}
    \widehat{\rho}_{B}^{\,k}(b) = \frac{1}{\sqrt{2\pi} \, |B|} \sum_{b' \in B}
    \exp \left(
    -\, \frac{{d^k(b,\, b')}^2}{2h^2}
    \right)
    \label{eq:KDE}
  \end{align}
\end{definition}

\noindent
Intuitively, the kernel density estimate of $b$ is inversely proportional to the distances to other batches, with diminishing weights for more distant ones.
The bandwidth $h$, the ``smoothing factor'', determines how quickly this drop-off occurs.

\paragraph{Density aggregation}
\name considers the prediction and output space as equally important.
Thus, we aggregate densities using the minimum: we regard the batch as rare if its density is low in either the prediction or output space:
\begin{align}
  \widehat{\rho}_{B}(b) = \min \left(
  \widehat{\rho}_{B}^{\,pred}(b),\,
  \widehat{\rho}_{B}^{\,out}(b)
  \right)
  \label{eq:minimum_density}
\end{align}
\noindent
If necessary, this approach can be generalized to fewer or more densities:
for unsupervised learning without ground truth, \name can only consider $\widehat{\rho}^{\,pred}$.
For models with multiple outputs and thus multiple predictions, we can compute the minimum in \cref{eq:minimum_density} across all respective densities.

\paragraph{Alternative distances}
We leverage an information-based distance metric to focus on the tail.
However, the density estimation is not tied to this method.
\name is flexible and can operate with any other provided distance metric.

\subsection{Sample selection}
\label{ssec:selection}

\name optimizes the sample-space coverage by iteratively discarding high-density batches.
It computes the densities for all samples in $S^* = S_{new} \cup S_{mem}$, randomly discards batches with a density-dependent probability~(\cref{alg:coverage}, \cref{line:while}--\ref{line:update}) until $|S^*| \leq C$, and updates densities after each discard.

Intuitively, \name assigns a high discard probability to batches with high density---batches for which many similar samples exist---thereby protecting rare low-density samples.
However, because noisy samples also seem ``rare,'' we must retain some probability of discarding rare samples.
\name achieves this by mapping densities to probabilities using softmax with temperature scaling~\cite{guoCalibrationModernNeural2017}:

\pagebreak

\begin{align}
  p^{discard}= \softmax \left(
  {\widehat{\rho}_B}/{T}
  \right)
  \label{eq:softmax}
\end{align}
where $\widehat{\rho}_B$ is a vector of densities for all batches $b\in B$, and $p^{discard}$ is a corresponding vector of discard probabilities.

The temperature $T$ allows balancing tail-focus with noise rejection.
A low temperature assigns a higher discard probability to the highest-density batch(es).
Conversely, a high temperature assigns more uniform discard probabilities, increasing the probability of discarding low-density batches.
At the extreme, the discard probability with $T \to 0$ is a point mass;
if \name is configured with T=0, we thus deterministically discard the highest-density batch. With $T \to \infty$, the probability becomes uniformly random.

\subsection{Retraining decision}
\label{ssec:changes}

Intuitively, retraining is beneficial if we collect \emph{new} information, \ie samples in areas of the sample space that were previously not covered.
That is, we should retrain only when the coverage of sample space increases.
\name's density estimation allows estimating this increase in information to guide the retraining decision.

\begin{definition}[Coverage increase]
  Let $B$ be a set of batches with density estimates $\widehat{\rho}_{B}$.
  We can approximate the region of the sample space covered by the samples in $B$:
  \begin{align}
    Coverage(B) = \sum_{b \in B} \widehat{\rho}_{B}(b)
  \end{align}
  Let $B'$ be a second set of sample batches.
  We can approximate the coverage increase, short $CI$, of $B$ with respect to $B'$, \ie the region of sample space covered by $B$ but not by $B'$:
  \begin{align}
    CI(B,\, B') = \sum_{b \in B} \min\left( \widehat{\rho}_{B}(b) - \widehat{\rho}_{B'}(b),\, 0 \right)
  \end{align}
  The \emph{relative} coverage increase $RCI$ from $B'$ to $B$ is then
  \begin{align}
    RCI(B,\, B') = CI(B,\, B') / Coverage(B)
  \end{align}
  where $RCI(B,\, B') \in [0, 1]$; $0$ means that the same area of sample space is covered, while $1$ indicates that $B$ covers an entirely different region of the sample space than $B'$.
\end{definition}

Hence, with $B$ the current memory batches and $B'$ those used for the last model training, $RCI(B,\, B')$ estimates the coverage increase since the last training. If it exceeds the user-defined threshold $\tau$, \name triggers retraining.
This approach presents a rational trade-off: the larger $\tau$, the longer we wait for changes to accumulate before retraining. With a small $\tau$, the model compensates for changes quicker at the cost of more retraining.
\name's training decision is sample-space-aware, it gives a rational argument that retraining is likely to be beneficial (even if there is no guarantee).

\section{Evaluation: Real-world benefits}
\label{sec:puffer}

We use Puffer~\cite{Puffer} to evaluate \name's real-world benefits.
This experiment aims to show that \name improves the tail performance of \emph{existing models} reliably without significantly impacting the average.
Puffer provides both a public dataset with data collected daily over several years and a publicly available model that we can retrain with \name and compare against the original.
This makes Puffer a perfect case study to investigate the following questions:

\begin{description}[nosep, topsep=1mm,itemsep=1.45mm]
  \item[Q1] \emph{Does} \name \emph{improve the tail predictions?}
    \hfill \textbf{\emph{Yes}}\\
    Over years of live and replay data,
    \name significantly improves the 1st percentile prediction score.

  \item[Q2]
    \emph{Does it improve the application performance?}
    \hfill \textbf{\emph{Yes}}\\
    On live Puffer, over \num{10} stream-years of data,
    \name achieves a \SI{14}{\percent} smaller fraction of stream-time spent stalled with only \SI{0.14}{\percent} degradation in image quality.

  \item[Q3] \emph{Does} \name \emph{avoid unnecessary retraining?}
    \hfill \textbf{\emph{Yes}}\\
    \name retrains 4 times in the first 8 days, and only 3 times in the following 9 months (7 times in total).

  \item[Q4]
    \emph{Are our improvements replicable?}
    \hfill \textbf{\emph{Most likely}}\\
    \name benefits appear replicable over different time periods of Puffer data.
    Moreover, its design parameters are intuitive and easy to tune.

  \item[Q5]
    \emph{Can} \name \emph{benefit existing solutions?}
    \hfill \textbf{\emph{Yes}}\\
    \name further improves tail predictions achieved by more advanced training or prediction strategies.

\end{description}%

\subsection{The Puffer project}
\label{ssec:puffer_overview}

The Puffer project is an ongoing experiment comparing ABR algorithms for video streaming~\cite{Puffer}.
Puffer streams live TV with a random assignment of ABR algorithms and collects Quality-of-Experience (QoE) metrics:
the mean image quality measured in SSIM~\cite{zhouwangImageQualityAssessment2004}
and the time spent with stalled video.

\fugu is the ABR algorithm proposed by Puffer's authors; it features a classical control loop built around a \emph{Transmission Time Predictor} (TTP), a neural network predicting the probabilities for a set of discretized transmission times.
The TTP was retrained daily with \meganum{1} random samples drawn from the past \num{2} weeks: \kilonum{\sim130} samples from the latest day and \SI{\sim10}{\percent} fewer samples for each previous day.
\fugufeb is a static variant, trained in February 2019 and never retrained.

\fugu was discontinued only \num{17} days after \name's current deployment. Hence, we can only compare \name's long-term performance to \fugufeb.
As discussed in~\cref{sec:introduction}, \fugu and \fugufeb achieve similar performance: Over almost three years, \fugu showed an SSIM improvement of \SI{0.17}{\percent} and a reduction in the time spent stalled of \SI{4.17}{\percent}.
Thus, improvements over \fugufeb would likely translate to similar---yet slightly lesser---improvements over the daily-retrained \fugu.

\begin{figure*}
  \begin{subfigure}[b]{0.24\textwidth}
    \centering
    \includegraphics[width=\textwidth,height=\textwidth]{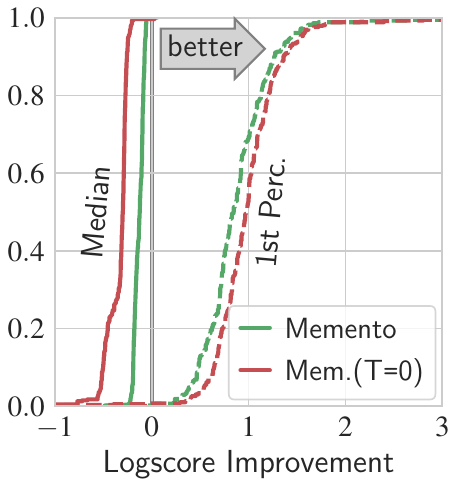}
    \caption{\name improves the prediction quality at the tail, with a relatively small impact on the median prediction quality.
    }
    \label{fig:score_improvement}
  \end{subfigure}\hfill%
  \begin{subfigure}[b]{0.24\textwidth}
    \centering
    \includegraphics[width=\textwidth,height=0.98\textwidth]{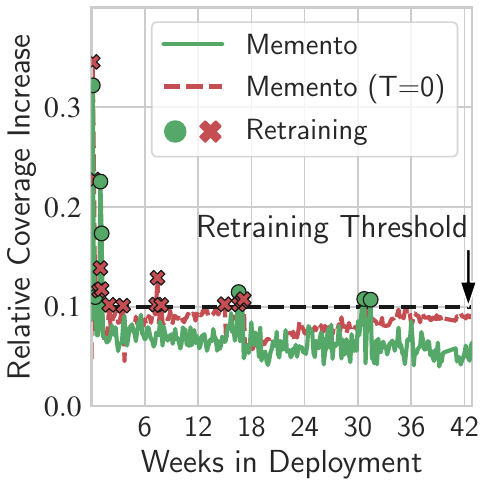}
    \caption{%
      \name avoids unnecessary retraining. Its deterministic variant ($T = 0$) is less efficient and does not converge.
    }
    \label{fig:retraining}
  \end{subfigure}\hfill%
  \begin{subfigure}[b]{0.24\textwidth}
    \centering
    \includegraphics[width=\textwidth,height=\textwidth]{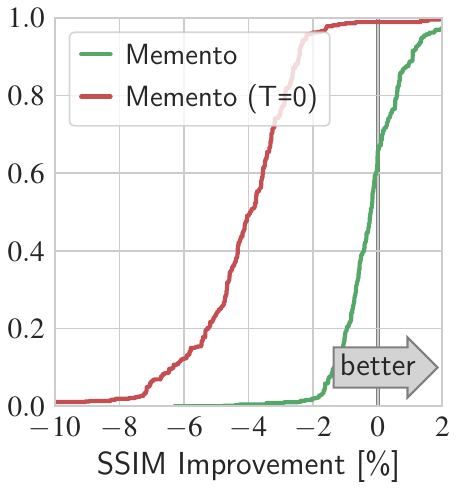}
    \caption{
      \name degrades the SSIM slightly (\SI{0,14}{\percent} worse).
      The det. variant degrades it notably (\SI{4,4}{\percent} worse).
    }
    \label{fig:qoe_ssim}
  \end{subfigure}\hfill%
  \begin{subfigure}[b]{0.24\textwidth}
    \centering
    \includegraphics[width=\textwidth,height=\textwidth]{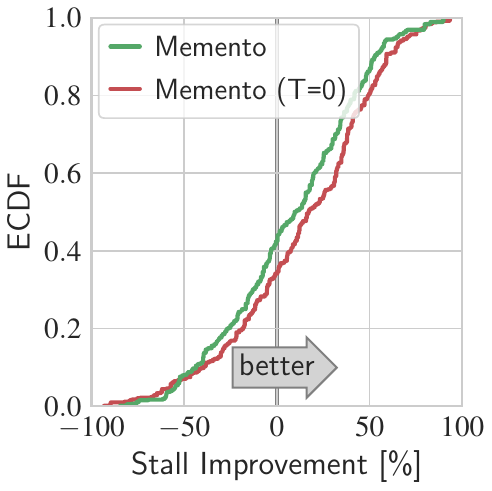}
    \caption{
      \name reduces the fraction of stream-time spent stalled by \SI{14}{\percent}, although results vary by day.
    }
    \label{fig:qoe_stalls}
  \end{subfigure}
  \caption{
    \name achieves its goal: it improves the tail prediction quality with minimal impact on the average (\cref{fig:score_improvement}).
    This requires little retraining (\cref{fig:retraining}) and translates into notable QoE improvements~(\cref{fig:qoe_ssim,fig:qoe_stalls}).
  }
  \label{fig:puffer_results}
\end{figure*}

\subsection{Retraining with \name}
\label{ssec:puffer_results}

We use \name to retrain \fugu's TTP:
every day, we use \name to select the training samples and decide whether to retrain.
We assess \name's benefits in two experiments:

\begin{description}[nosep, topsep=1mm,itemsep=1mm]
  \item[\emph{Deployment}]
    We deploy on Puffer two \name variants, \ie two variants of \fugu using \name for retraining the TTP.
    One uses \name's default parameters (see below), and the other deterministic sample selection (\ie using temperature $T=0$, \cref{ssec:selection}).
    We collected data over \num{292} days (from Oct. 2022 to Jul. 2023), totaling around \num{10.8} stream-years of video data per variant.
    This experiment allows answering \textbf{Q1}, \textbf{Q2}, and \textbf{Q3}.
  \item[\emph{Replay}]
    To confirm the deployment observations, evaluate design choices, and benchmark the impact of \name's parameters, we replay Puffer data collected since 2021.
    To reduce the bias from a particular starting day, we replay \num{3} instances with \num{6} months of video data each and a total of \num{90} stream-years of video data.
    This experiment allows answering \textbf{Q4} and \textbf{Q5}.
\end{description}

\paragraph{Metrics}
We access \name along three dimensions:
\begin{itemize}[nosep, topsep=1mm,itemsep=1mm]
  \item \emph{Prediction quality} is measured with the logarithmic score
        ${logscore(y) = \log p(y)}$~\cite{gneitingStrictlyProperScoring2007}; $y$ is the video chunk transmission time and $p(y)$ the TTP-predicted probability.
        Score improvement equals TTP loss decrease.%
        \footnote{%
          The logarithmic score is a commonly used metric for probabilistic predictions~\cite{gneitingStrictlyProperScoring2007} like those produced by the Puffer TTP model.
          It is closely related to the \emph{cross entropy loss} used to train the TTP: this is also known as the \emph{logarithmic loss} and is the negative logarithmic score.
        }
        Available in deployment and replay.
  \item \emph{Application performance} is measured with user QoE;
        Only available for the deployment experiment, where real user streams are impacted by the predictions and the resulting QoE can be measured.
  \item \emph{Training resource utilization} is measured by the number of retraining events in deployment and replay.
\end{itemize}

\paragraph{Parameters}
We set \name's memory capacity $C$ to \meganum{1} samples (same as the original \fugu model).
Default parameters are a retraining threshold $\tau$ of \num{0.1}, batching size $b$ of \num{256}, kernel bandwidth $h$ of \num{0.1}, and temperature $T$ of \num{0.01}.
We evaluate the impact of different parameter values in \cref{ssec:benchmarks}.

\subsection{Deployment results}
\label{ssec:deployment}

In this section, we show that \name effectively improves the tail prediction quality with little retraining and that this translates into QoE improvements over \fugufeb.

In \cref{app:puffer_comparison_extra}, we provide additional plots with the evolution of QoE and predictions over time, and aggregate plots like those published on the Puffer paper and website~\cite{yanLearningSituRandomized2020, Puffer}.\footnote{%
  Our results per algorithm differ from official Puffer plots, as Puffer excludes some sessions in an attempt to exclude effects such as ``client decoder too slow,'' while we consider all data points.
  As the filtering is ABR-independent, it does not impact relative results between ABRs.
}

\paragraph{Prediction quality}
\cref{fig:score_improvement} shows the score difference ECFD between \name and \fugufeb in median and 1st-percentile scores on the whole deployment (higher is better).

We observe that \name improves tail predicition performance as intended, with a slight---yet expected---degradation on average:
as memory is finite and \name purposefully prioritizes ``rare'' samples, it must remove samples for the most common cases.
The deterministic version of \name prioritizes the tail more aggressively, which leads to slightly better tail improvements and worse median degradation.

\paragraph{Retraining count}
\cref{fig:retraining} shows the relative coverage increase $RCI$ between the current memory and the one last used to retrain the model, as well as the retraining threshold $\tau=0.1$.
We observe about eight ``warm-up'' days where \name retrains four times.
Afterward, RCI remains low and retraining due to changes in the data (RCI peaks) happens three times.
This shows that there are fewer ``new patterns'' to learn from over time; retraining daily is unnecessary.

Conversely, the deterministic variant of \name keeps accumulating samples, exhibiting a different RCI pattern.
After training five times during ``warm-up,'' it trained \num{9} more times.
Where RCI for default Memento stabilizes, the RCI of the deterministic variant slowly but steadily increases over time while it accumulates rarer and rarer samples, as it does not reject noise.
Essentially, this variant ``never forgets.''

\paragraph{Application performance}
\cref{fig:qoe_ssim,fig:qoe_stalls} show the relative QoE improvements achieved over \fugufeb for the SSIM and the time stalled (higher is better), respectively.%
\footnote{
To avoid bias towards either \name or \fugu, we show the symmetric percent difference using the maximum:
$100\cdot{(x - y)}/{\max (x,\ y)}$.
}

First, we observe that \name only marginally affects the SSIM; the average SSIM is \num{17.12} and \num{17.14} for \name and \fugufeb, resp. (not shown).
\name's deterministic variant affects the SSIM more; its average is \num{16.39} (not shown) and the SSIM is consistently worse than \fugufeb (\cref{fig:qoe_ssim}).

Second, \cref{fig:qoe_stalls} shows that the improvement in stalls compared to \fugufeb is almost the same for both variants of \name, even though there are large day-to-day variations:
some days, \fugufeb stalls much less than \name, and vice versa.
Over the entire \num{292} days of deployment, \name spent a fraction of \SI{0.2}{\percent} of stream-time stalled, compared to \SI{0.24}{\percent} for \fugufeb.
In relative terms, \name spent a \SI{14}{\percent} smaller fraction of stream time stalled than \fugufeb.

In the results above, we already see that \name performs slightly worse without noise rejection (\ie with $T=0$).
In a previous deployment, we observed that never forgetting ultimately prevented enough average samples from remaining in memory, which destroyed the average performance~(\cref{app:puffer_comparison_extra}).
The latest version of \name made the deterministic variant more robust but we see the signs of noise accumulation (worse predictive performance, more frequent retraining, steadily rising RCI).
By contrast, the probabilistic default \name naturally forgets noise and stabilizes, as can be seen in the RCI in \cref{fig:retraining}.

Finally, we observe that \name reduces stalls by \num{3.5} times as much as retraining daily with random samples.
In \cref{app:puffer_comparison_extra}, \cref{fig:appendix_past_scores} we show \cref{fig:score_improvement} overlaid with the score improvements of \fugu in the past.\footnote{%
  \cref{app:puffer_comparison_extra}, \cref{fig:appendix_past_scores} must be considered with caution, as the underlying data comes from different time periods and may not be comparable.
}
We observe that random retraining improved the tail prediction scores significantly less and even worsened them on \SI{20}{\percent} of days.

\begin{figure*}
  \begin{subfigure}[t]{0.32\textwidth}
    \centering
    \includegraphics[scale=0.5]{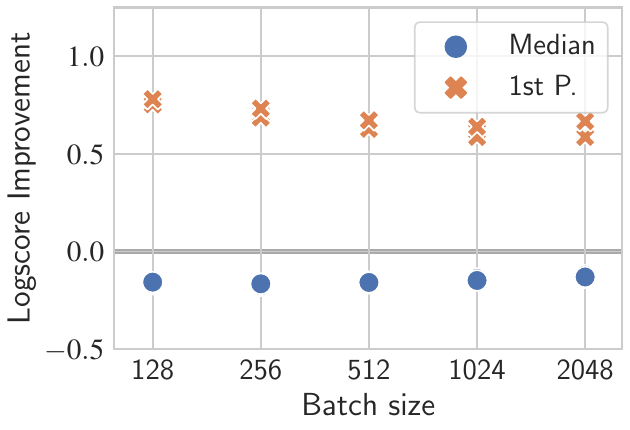}
    \caption{%
      Larger batches slightly degrade the tail with no impact on the median.
    }
    \label{fig:batchsize}
  \end{subfigure}\hfill%
  \begin{subfigure}[t]{0.32\textwidth}
    \centering
    \includegraphics[scale=0.5]{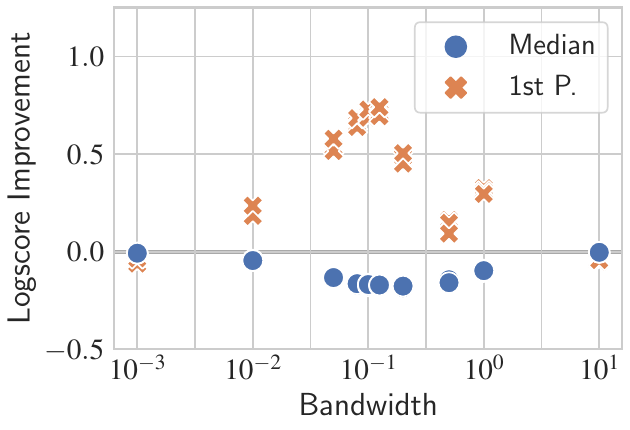}
    \caption{%
      A kernel bandwidth around \num{0.1} is best.
      Extreme values nullify benefits.
    }
    \label{fig:bandwidth}
  \end{subfigure}\hfill%
  \begin{subfigure}[t]{0.32\textwidth}
    \centering
    \includegraphics[scale=0.5]{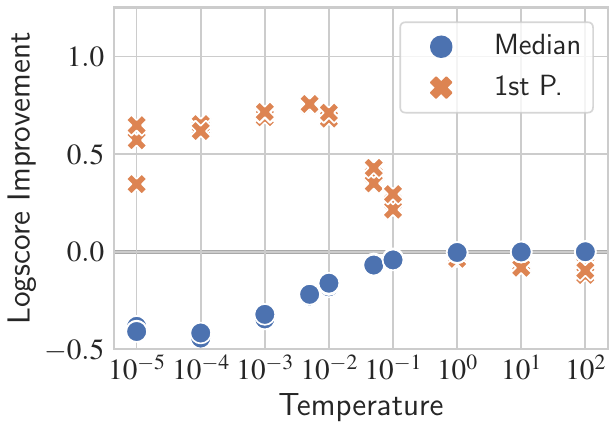}
    \caption{
      $T = 0.01$ is a good trade-off between tail prioritization and randomness.
    }
    \label{fig:temperature}
  \end{subfigure}
  \vspace{-1mm}  %
  \caption{Each marker shows the median prediction improvement of \name over \fugu (random selection) over a \num{6}-month replay, measured as median and 1st perc. improvement each day. Results are consistent across replays.}
  \label{fig:benchmarks}
  \vspace{-1mm}  %
\end{figure*}

\paragraph{From predictions to QoE}
One may wonder why the average prediction degradation~(\cref{fig:score_improvement}) does not seem to strongly impact image quality~(\cref{fig:qoe_ssim}), and, conversely, why significant tail improvements yield only a modest reduction in stalls~(\cref{fig:qoe_stalls}).
Our results illustrate the complex relationship between predictions and QoE, including the closed-loop control logic, which uses the predictions and aims to keep the video buffer at the receiver sufficiently full to avoid stalling.

Looking closer, we noticed that the transmission time of most chunks is very small, and most prediction errors are also small time-wise.
Hence, slightly worse predictions have little effect on the buffer fill level; the controller has time to compensate and maintain image quality.
Moreover, since most prediction errors overestimate the transmission time (not shown), it makes the closed-loop control more conservative. Thus, it manages to keep stalls low but struggles to maintain high image quality (compare \cref{fig:qoe_ssim,fig:qoe_stalls}).
Further investigations of the interplay between prediction quality and application performance would be interesting but are beyond the scope of this work.
To facilitate further research, we publish all our retrained models including their training sample selection alongside the Puffer QoE data.

\subsection{Replay results}
\label{ssec:benchmarks}
In this section, we confirm that \name{}'s benefits are replicable and not just an artifact from deploying at an ``easy time,'' its design decisions are justified, and it complements existing techniques.
To do this, we replay \num{3} non-overlapping instances of \num{6} months with \numlist{25;39;26} stream-years of video-data respectively and \SI{3}{M} samples on average per day.

To monitor the memory quality over time, we disable threshold-based retraining and retrain every \num{7} days; one must retrain and test the model to assess whether the right samples were selected.
We evaluate each day in terms of prediction improvement over retraining with random samples (\fugu) and report the mean over each \num{6}-month period.
We show the entire time series for each experiment in ~\cref{app:puffer_comparison_extra}.

\paragraph{Replicability}
All plots in \cref{fig:benchmarks} show three data points per setting, which are average performance numbers over the entire \num{6} month period. We observe that all results are fairly stable,
which gives reasonable confidence about the replicability of \name's benefits on this use case.

\paragraph{Batch size}
\cref{fig:batchsize} shows \name's prediction performance over the batch size;
larger sizes improve scalability but make sample selection more coarse-grained,
which should hurt performance~(\cref{ssec:distances}).
We observe a slight tail performance drop for larger batch sizes with little change on average.

Regarding scalability, differences are more pronounced:
using a single CPU core to process \meganum{4} samples takes on average
\SI{200}{\second} with a batch size of \num{128},
\SI{48}{\second} with \num{256} (the default), and
\SI{7}{\second} with \num{1024}.
Benefits flatten out for larger batch sizes. Overall, computation is dominated by distance computation; batching the samples takes only about \SI{2.5}{\second}.

\paragraph{Bandwidth}
For each batch, \name estimates how close nearby batches are; the kernel bandwidth $h$ determines what ``nearby'' means~(\cref{ssec:density}).
As the computed distances $JSD(P, Q) \in [0, 1]$, bandwidths $> 1$ over-smooth (all batches are always ``nearby''), and bandwidths $\ll 1$ under-smooth (no other batches are ever ``nearby''). Both cases nullify the idea of estimating density, effectively making the sample selection random.
\cref{fig:bandwidth} confirms this intuition: at the extremes, \name performs like a random selection.
We obtain the best tail improvement with a bandwidth around \num{1e-1}.

\paragraph{Temperature}
\cref{fig:temperature} shows \name's prediction performance over the temperature $T$; a low temperature strongly prioritizes rare samples at the risk of accumulating noise, while a high temperature rejects noise by making the sample selection more random~(\cref{ssec:selection}).
As expected, a lower temperature yields better tail performance but degrades the average.
The trade-off is not linear, though; we can select a temperature that provides tail benefits with minimal impact on the average.
The best trade-off is a temperature around \num{1e-2}.

\begin{figure}[t]
  \centering
  \includegraphics[scale=0.5]{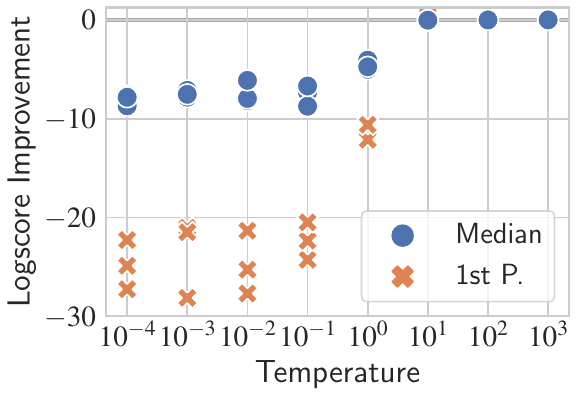}
  \vspace*{-1mm}
  \caption{%
    Loss-based selection is worse and less robust.
  }
  \vspace*{-1mm}
  \label{fig:alt_loss}
\end{figure}

\paragraph{Alternative selection metrics}
\cref{fig:alt_loss} shows the performance of using loss as an alternative selection metric: we still use temperature-based probabilistic selection but prefer to discard samples with a low loss rather than high density.
At best, it gives tail improvements about half as \name, but it is much harder to tune:
the benefits vanish for a slightly higher temperature.
With a lower temperature, \ie selecting more strongly based on loss, the model performance decreases drastically, which mirrors our observations in~\cref{sec:overview}.

We evaluate additional metrics in \cref{app:puffer_comparison_extra}: prediction confidence, label counts, and whether a sample belongs to a stalled session or not. In summary, these perform worse or equal to loss-based selection in the best case, and most of them are as sensitive to tune.
Probabilistic selection based on density performs better and is less sensitive (see~\cref{fig:temperature}).
We also show detailed results for Euclidean distances (\cref{ssec:distances}).

\paragraph{Alternative training decision}
We compare \name's retraining decision based on the \emph{relative coverage increase} RCI with a loss-based decision (not shown).
We observe that for samples selected by \name, either decision is effective.
Overall, a coverage-based decision provides greater control over retraining frequency but struggles with low thresholds (\eg \SI{5}{\percent}).
As \name is probabilistic, the estimated RCI fluctuates at each iteration, which can be observed in \cref{fig:retraining}, and the retraining threshold should be set above these fluctuations.
It may be possible to further improve \name by smoothing the RCI or by attempting to remove the random fluctuations.
We leave this challenge for future work.

\begin{figure}[h]
  \centering
  \includegraphics[scale=0.55]{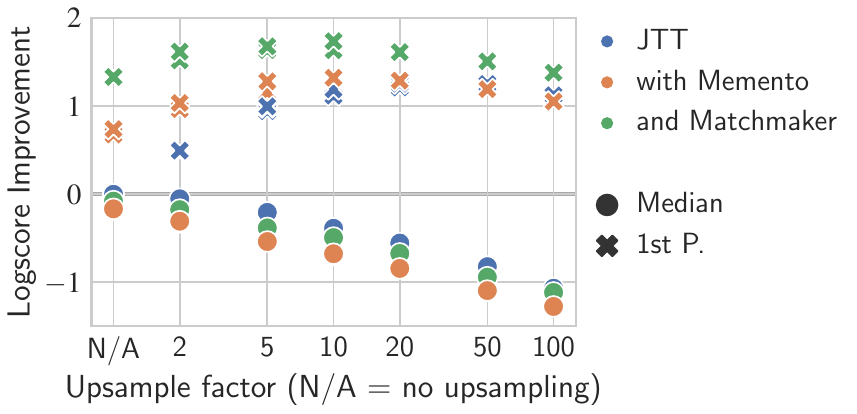}
  \caption{%
    \name complements training and prediction improvements such as JTT and Matchmaker.
  }
  \label{fig:combinations}
\end{figure}

\paragraph{\name $+$ Matchmaker $+$ JTT}
MatchMaker~\cite{mallickMatchmakerDataDrift2022} improves predictions by using an ensemble of models (default is \num{7}) combined with an online algorithm to select the best model to make a prediction for each sample.
However, this is limited by the performance of the individual models.
Using an ensemble of models trained with a random selection, even with an oracle choosing the best model, we can only improve tail performance by half as much as a single model trained with \name.
However, we can get the best of both worlds by using MatchMaker with an ensemble of \name-trained models, which yields double the tail performance with less decrease in median performance compared to a single \name-trained model~(`no upsampling' in \cref{fig:combinations}).

JTT~\cite{liuJustTrainTwice2021} improves performance by training twice: after the first training, misclassified samples are upsampled in the second and final training.
\cref{fig:combinations} shows the same performance tradeoffs for JTT and \name: both improve the tail and degrade the median:
JTT with an upsampling factor of \num{3} is roughly equivalent to \name's sample selection with `normal' training.
Yet this comes at different resource costs: JTT requires up to double the training time and resources, depending on how long the first training step is.
Training models like Puffer takes time in the range of hours~\cite{yanLearningSituRandomized2020} and often requires expensive hardware (\eg GPUs).
Memento is more resource efficient: even on a single CPU core, it can process millions of samples in a few minutes (see above).
However, JTT and \name are not in competition, but complementary.
We observe the best performance by combining \name-selected samples with JTT's training and observe even further improvements when the resulting models are used in a MatchMaker ensemble for predictions~(\cref{fig:combinations}).

\begin{figure}[h!]
  \centering
  \includegraphics[scale=0.5]{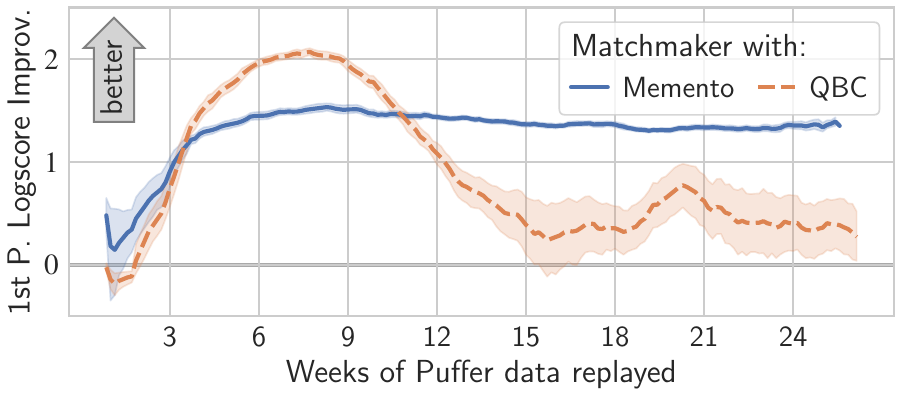}
  \caption{%
    Over time, \name outperforms QBC.
  }
  \label{fig:memento_vs_qbc}
\end{figure}

\paragraph{Query-by-Committee}
A \name-trained ensemble not only outperforms an ensemble trained with random samples but also an ensemble trained with the Query-by-Committee (QBC) algorithm~\cite{seungQueryCommittee1992} (\cref{fig:memento_vs_qbc}).
QBC selects samples with the highest prediction entropy between all members of the ensemble.
In theory, these samples contain the most information for learning.
We repeat the MatchMaker-Oracle experiment and compare an ensemble of \name-trained models with an ensemble of QBC-trained models.

\cref{fig:memento_vs_qbc} shows that the tail improvements of QBC initially exceed \name, and are comparable to \name with JTT (\cref{fig:combinations}).
However, the QBC ensemble degrades over time and ultimately settles on less than a third of \name's improvement.
We suspect this comes from QBC accumulating noise.
Both \name and QBC initially pick up noise (low density; high entropy).
\name's probabilistic approach prevents noise from accumulating (\cref{ssec:selection}) while QBC degrades:
selecting noise biases the model toward random predictions, increasing the prediction entropy of any sample, thus decreasing QBC's ability to identify informative samples.

\newcommand*{\random}{\textsf{Random}\xspace}
\newcommand*{\fifo}{\textsf{FIFO}\xspace}
\newcommand*{\lars}{\textsf{LARS}\xspace}

\begin{figure*}[t!]
  \begin{subfigure}[t]{0.32\textwidth}
    \centering
    \includegraphics[scale=0.5]{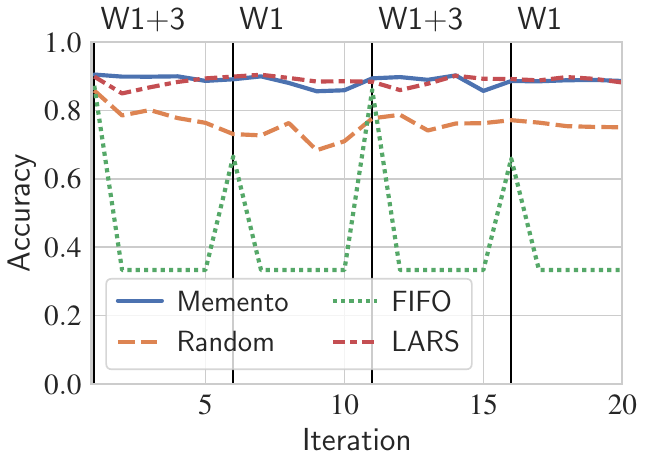}
    \caption{%
      Rare and infrequent traffic patterns:
      Only \lars and \name avoid forgetting.
      Traffic for W2 is always present; for W1 and W3 only where marked.
    }
    \label{fig:sim_periodic}
  \end{subfigure}\hfill
  \begin{subfigure}[t]{0.32\textwidth}
    \centering
    \includegraphics[scale=0.5]{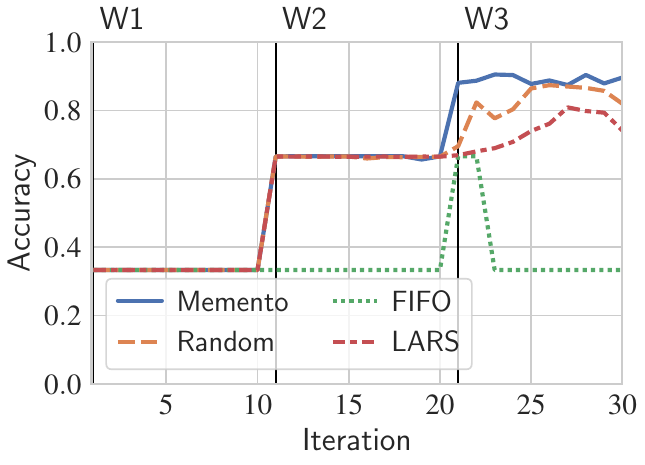}
    \caption{%
      Incremental learning:
      \name efficiently integrates workload traffic patterns as they appear sequentially where marked, replacing the previous ones.
    }
    \label{fig:sim_sequential}
  \end{subfigure}\hfill
  \begin{subfigure}[t]{0.32\textwidth}
    \centering
    \includegraphics[scale=0.5]{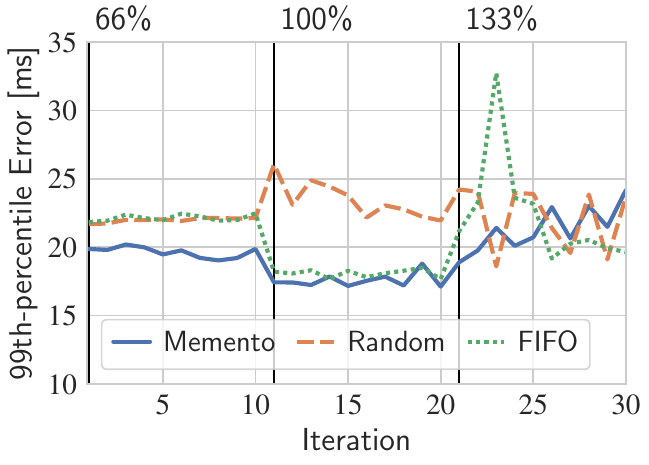}
    \caption{%
      Increasing congestion:
      \name has comparable or lower 99th percentile error
      for the marked level of network utilization,\ie increasing congestion.
    }
    \label{fig:sim_congestion}
  \end{subfigure}%
  \caption{%
    \name can handle various traffic patterns and prediction types, outperforming other approaches.
  }
\end{figure*}

\section{Evaluation: Synthetic shifts}
\label{sec:simulation}

In the previous section, we showed that \name provides significant benefits in a real-world use case.
In this section, we use {ns-3}~\cite{rileyNs3NetworkSimulator2010} simulations of data center workloads to show that sample selection with \name applies to other settings as well.
Specifically, we show that \name:

\begin{enumerate}[nosep, topsep=0.5mm,itemsep=1mm]
  \item ensures good tail performance by reliably prioritizing samples from infrequent traffic patterns (\cref{ssec:sim_tail_perf});

  \item picks up new patterns quickly (\cref{ssec:sim_no_forgetting});

  \item is applicable to classification and regression (\cref{ssec:sim_regression}).
\end{enumerate}

\subsection{Experimental setup}

\paragraph{Sample selection strategies}
We compare \name with two baselines: \random (random sampling) and \fifo (keep recent samples). In addition, we compare it to the state-of-the-art \lars (Loss-Aware Reservoir Sampling,~\cite{buzzegaRethinkingExperienceReplay2020}).
\lars uses several improvements to random sampling for classification, and has two stages: first, it randomly chooses to keep or discard a new sample, with probability exponentially decreasing over time; second, it considers both label counts and loss to decide which in-memory sample to replace.

\paragraph{Parameters}
For \name, we use the same default parameters as before: a batching size of \num{256}, kernel bandwidth $h$ of \num{0.1}, and temperature $T$ of \num{0.01}.
We reduce the memory capacity $C$ to \kilonum{20} samples (\ie \num{1}/\num{50} compared to \cref{sec:puffer}) for two reasons:
\first We aim to show the limitations of different sample selection strategies, which is easier to do using small memories;
\second \lars scales poorly, making comparing performance on larger memory sizes impractical---we optimized the original \lars implementation to scale to \kilonum{20} samples. Our optimization is available in our artifacts.

\paragraph{Workloads}
The simulation setup~(\cref{app:sim_extra}, \cref{fig:simulation_network}) consists of two nodes and applications sending messages
whose sizes follow three empirical traffic distributions from the Homa project (\cref{fig:simulation_distributions},~\cite{montazeriHomaReceiverdrivenLowlatency2018}): Facebook web server (\textbf{W1}), DC-TCP (\textbf{W2}), and Facebook Hadoop (\textbf{W3}). W1 and W3 are similar size-wise, while W2 messages are about an order of magnitude smaller.
For each workload, we generate \num{20} traffic traces of \SI{1}{\minute} each.
During this time several senders transmit a combined \SI{20}{Mbps}, resulting in an average network utilization of \SI{66}{\percent}.
We repeat this process by injecting additional cross traffic to reach an average utilization of \SI{100}{\percent} and \SI{133}{\percent}, respectively.
We use different random initializations to generate a total of 180 distinct runs that we combine in various iterations in the following experiments.

\paragraph{Models}
We compare the selection strategies for two neural networks; one for classification and one for regression.
The classification model predicts the application workload: Each input is a trace of the past \num{128} application packet sizes, and the model predicts the probabilities for each workload.
The regression model predicts the next transmission time from past packets: Each input contains a trace of the past \num{127} packet sizes and transmission times, the current packet size, and the model predicts the transmission time.
See \cref{app:sim_extra} for architecture and hyperparameter details.

\paragraph{Metrics}
For classification tasks, we measure the balanced accuracy, \ie the accuracy obtained over an equal number of evaluation samples per workload, ensuring equal importance of each workload.
In other words, performance is evaluated over an equal distribution of overall workloads, regardless of whether they are present at the current iteration.
Good performance requires both picking up new patterns quickly and avoiding catastrophic forgetting.

For regression, we investigate changes in traffic distribution (see below) and measure the 99th percentile absolute prediction error over the data in the latest iteration.
Good performance requires picking up new patterns quickly.

\subsection{Classification: Rare patterns}
\label{ssec:sim_tail_perf}

In the first experiment, we show that Memento successfully picks up samples from infrequent traffic patterns. To do so, we use highly imbalanced traffic: W1 and W3 only constitute \SI{\leq 2}{\percent} of overall traffic.
Good tail performance implies high accuracy not only for W2 but also for W1 \& W3.

\paragraph{Setup}
We use the classification model and iterate over samples from \num{20} runs.
We use W2 at every iteration, representing a large part of traffic that remains relatively unchanged.
On top of that, we include W1 once every five iterations, and W3 once every ten iterations; they represent sporadic traffic patterns
that make up for \SIlist{1.3;0.5}{\percent} of traffic respectively.

\paragraph{Results}
\name and \lars retain sufficient samples from each workload and show the best accuracy over all iterations~(\cref{fig:sim_periodic}).
On the other hand, \fifo shows good accuracy only while all workloads are present, as the large number of samples of W2 quickly overwrites W1 \& W3 otherwise.
While \random achieves better results than \fifo, it ultimately retains too few samples of W1 \& W3, as they make of less than \SI{2}{\percent} of samples in memory.

\subsection{Classification: Incremental learning}
\label{ssec:sim_no_forgetting}

Next, we show that \name quickly picks up new patterns and avoids catastrophic forgetting in an `Incremental Learning' setting, which is known for its challenging nature~\cite{farquharRobustEvaluationsContinual2019}.

\paragraph{Setup}
We use the same model and setup as \cref{ssec:sim_tail_perf}, but iterate over samples from each workload sequentially;
first W1, then W2, and finally W3, for \num{10} iterations each.

\paragraph{Results}
We find that overall, \name exhibits the best performance. Both \random and \lars struggle because of their sample selection rate~(\cref{fig:sim_sequential}); these two flavors of random memory
avoid forgetting by decreasing the probability of selecting new samples over time.
When W3 is introduced, they are slow to incorporate new samples~(\cref{app:sim_extra}, \cref{table:sim_end_fraction}).
While both \lars and \random are slow to react, the fact that the balanced accuracy of \lars is worse than \random's is mostly an artifact of the similarity of W1 and W3: \lars has (desirably) retained more samples of W1,
yet this causes its model to mistake W3 for W1 more often than \random,
which has forgotten most of W1 and is consequently less biased.
By manually tuning the sampling rate of \lars to be much more aggressive, we were able to achieve the same performance as \name (not shown).
This highlights the benefit of the self-adapting nature of \name's sample-space-aware approach:
If a new label appears, \name discovers that this part of the sample space is not well covered yet.
It quickly prioritizes discarding common in-memory samples to retain samples from the new label.

\subsection{Regression}
\label{ssec:sim_regression}

In this experiment, we show that \name is applicable to regression and handles complex traffic changes.
We iterate from \num{66} to \SI{133}{\percent} network utilization, which presents more complex gradual changes in traffic patterns than the abrupt changes in workload distributions~(\cref{ssec:sim_no_forgetting,ssec:sim_tail_perf}).

\paragraph{Setup}
We iterate over traffic from all workloads using runs with increasing congestion.
For the first \num{10} iterations, we use runs with \SI{66}{\percent} network utilization, followed by \num{10} iterations of \SI{100}{\percent}, and finally \num{10} iterations of \SI{133}{\percent}.
We report the 99th percentile error under the current traffic conditions.

\paragraph{Results}
\name generally shows the lowest 99th percentile prediction error~(\cref{fig:sim_congestion}).
\random is slow to react to the new patterns and requires several iterations to adjust to the new traffic conditions.
Perhaps surprisingly, \fifo performs well up to \SI{100}{\percent} utilization but shows very unstable performance for \SI{133}{\percent}.
\lars is not applicable to regression.

\section{Related work \& Discussion}
\label{sec:related}

\paragraph{Limitations}
Let us address the elephant in the room: \name cannot do anything with a bad model or dataset. It helps identify the most useful samples for training, but those samples must be present in the dataset in the first place, and the model must be capable of learning from them.

We find that density-based selection performs well but is probably not optimal. It addresses the problem of dataset imbalance well, but it is less effective to differentiate ``hard-to-learn'' from ``easy-to-learn'' samples, which is better captured by the model loss.
Combining both would likely be beneficial.

Finally, density computations limit \name's scalability. Batching helps~(\cref{fig:distance_comparison}), but it would not be enough to process data streams with billions of samples per day.

\paragraph{Continual learning}
Continual learning and dataset imbalance correction are well-studied problems.
Fundamentally, continual learning suffers from the \emph{stability-plasticity} dilemma~\cite{ditzlerLearningNonstationaryEnvironments2015}:
a stable memory consolidates existing information yet fails to adapt to changes, while a plastic memory readily integrates new information at the cost of forgetting old information.
Forgetting old-yet-still-useful information is known as \emph{catastrophic forgetting}~\cite{mcclellandWhyThereAre1995}.
Continual learning approaches aim to be as plastic as possible while minimizing catastrophic forgetting.
They can be broadly categorized as either prior-based or rehearsal-based~\cite{buzzegaRethinkingExperienceReplay2020}.
Prior-based methods aim to prevent catastrophic forgetting by protecting model parameters from later updates~\cite{kirkpatrickOvercomingCatastrophicForgetting2017,parisiContinualLifelongLearning2019,zenkeContinualLearningSynaptic2017}.
Rehearsal-based methods collect samples over time in a replay memory and aim to prevent forgetting by learning from both new and replayed old data~\cite{buzzegaRethinkingExperienceReplay2020,iseleSelectiveExperienceReplay2018,ratcliffConnectionistModelsRecognition1990}.
Hybrid methods combine both, \eg training with a replay memory and a loss term penalizing performance degradation on old samples~\cite{rolnickExperienceReplayContinual2019}.

\name builds on previous rehearsal-based approaches, incorporating ideas such as coverage maximization~\cite{debruinImprovedDeepReinforcement2016}.
It extends existing ideas by considering both prediction and output spaces, leveraging temperature scaling to control the tail-focus and introducing a novel \emph{coverage increase} criterion to reason about when to retrain.
Similar to \name, Arzani et al.~\cite{arzaniInterpretableFeedbackAutoML2021} also suggest using the ``right samples'' to retrain AutoML systems~\cite{heAutoMLSurveyStateoftheart2021}: They use the disagreement among a set of models to identify ``the tail'' and guide the user to collect more tail samples and add them to the training set.
However, this does not apply to all networking applications; \eg we cannot ``force'' streaming sessions to come from rare network paths or experience particular congestion patterns over the Internet; we must do with the available samples.

\paragraph{Distribution shift detection}
Continual learning closely relates to a branch of research aiming to keep models up-to-date upon changes in the data-generating process---known as distribution shifts.
State-of-the-art methods rely on statistical hypothesis testing~\cite{hinderNonParametricDriftDetection2020} or changes in empirical loss~\cite{tahmasbiDriftSurfStableStateReactiveState2021}, plus a time-based window (or multiple parallel windows)~\cite{ditzlerLearningNonstationaryEnvironments2015}.
When a change is detected, these algorithms advance the time window(s), discard ``outdated'' samples whose timestamp falls outside of the window(s), and retrain.

Shift detection algorithms make \emph{sample-space-aware} retraining decisions but lack a comparable selection strategy.
Once a change is detected, they discard all old samples.
This approach is too coarse-grained for networking: Network traffic is composed of many patterns~(\cref{app:puffer_comparison_extra}, \cref{fig:appendix_batch_coverage}) and not all patterns get outdated at once.

Matchmaker~\cite{mallickMatchmakerDataDrift2022} proposes a more incremental approach to shifts in networking data.
It uses an ensemble of models and ``matches'' each sample to the model trained with the most similar data.
However, it does not address the sample selection problem:
If no model in the ensemble is good at the tail, matching will not help.
By contrast, \name builds upon ideas originating from shift detection~(BBDR~\cite{liptonDetectingCorrectingLabel2018}) and extends the sample-space-aware strategy to the sample selection to train better models.
Our evaluation shows that a single \name-trained can outperform Matchmaker at the tail, even if using an oracle to match the optimal model~(\cref{ssec:benchmarks}).

\paragraph{Experience replay for reinforcement learning}
While we evaluate \name in the context of supervised learning, it may also be used for \emph{reinforcement learning} (RL), which is popular approach to ML-based ABR~\cite{xiaGenetAutomaticCurriculum2022,maoNeuralAdaptiveVideo2017}.
In fact, RL commonly uses a replay memory~\cite{vanhasseltDeepReinforcementLearning2016,riedmillerNeuralFittedIteration2005},
and it has been shown that it can greatly improve RL performance~\cite{aljundiOnlineContinualLearning2019,rolnickExperienceReplayContinual2019}.

\paragraph{Transformers and other large models}
In recent years, large models with billions of parameters have become popular in natural language processing~\cite{devlinBERTPretrainingDeep2019,openaiGPT4TechnicalReport2023,touvronLlamaOpenFoundation2023} and computer vision~\cite{hanSurveyVisionTransformer2022,khanTransformersVisionSurvey2021}, usually based on Transformer architectures~\cite{vaswaniAttentionAllYou2017};
there are also first trials in networking~\cite{dietmullerNewHopeNetwork2022,leRethinkingDatadrivenNetworking2022a,mondalWhatLLMsNeed2023}.

We have evaluated \name primarily on small models to investigate to impact of a \emph{smarter} sample selection instead of using more resources such as more complex models and cannot claim with confidence that we would see comparable improvements for them.
However, there is mounting evidence that even large transformer models suffer from data bias~\cite{birhaneLAIONsInvestigatingHate2023,longprePretrainerGuideTraining2023,zhaoUnderstandingEvaluatingRacial2021}, which \name may help to address.

Furthermore, an important part of these models is an encoder that translates text, images, or other data into a latent space.
Latent space representations have seen use in clustering~\cite{mukherjeeClusterGANLatentSpace2019}, and \name may benefit from computing sample distances in this latent space, similar to how it uses BBDR.

\paragraph{Data processing}
Data validation and augmentation are important steps of any ML pipeline.
Especially in ML systems that are evolving over time, new bugs may be introduced any time the model or data collecting system are updated.
These bugs may lead to erroneous data, and data validation is necessary to prevent such data from becoming part of the training data~\cite{polyzotisDataValidationMachine2019}.
Furthermore, collected data is often augmented to address dataset imbalance by generating additional synthetic samples or upscaling existing ones, which can improve performance and reduce overfitting~\cite{wongUnderstandingDataAugmentation2016}.
One example is Just Train Twice (JTT)~\cite{liuJustTrainTwice2021}, which trains a model, uses it to identify misclassified samples, upsamples those, and trains again.
As a replay memory, \name operates between the validation and augmentation steps, and complements them.
For example, we showed in \cref{ssec:benchmarks} that JTT's upsampling is more effective when \name has identified tail samples first.
Only validated data should be considered by the sample selection, and selected samples may be augmented.
In particular, the memory should only store non-augmented samples, as augmenting the data before passing it to the memory can result in the sample selection to overfit to the augmentation~\cite{buzzegaRethinkingExperienceReplay2020}.

\paragraph{Generalization}
\name could be applied to other ML-based networking applications, including congestion control~\cite{abbaslooClassicMeetsModern2020,jayDeepReinforcementLearning2019,nieDynamicTCPInitial2019,winsteinTCPExMachina2013},
traffic optimization~\cite{chenAuTOScalingDeep2018},
routing~\cite{valadarskyLearningRoute2017},
flow size prediction~\cite{dukicAdvanceKnowledgeFlow2019,poupartOnlineFlowSize2016},
MAC protocol optimization~\cite{jogOneProtocolRule2021,yuDeepReinforcementLearningMultiple2019},
traffic classification~\cite{busse-grawitzPForestInNetworkInference2019,wichtlhuberIXPScrubberLearning2022},
network simulation~\cite{zhangMimicNetFastPerformance2021}, or DDoS detection~\cite{wichtlhuberIXPScrubberLearning2022}.
Networking has proven to be a challenging environment for ML, and many proposed systems have only delivered modest or inconsistent improvements in real networks~\cite{yanLearningSituRandomized2020,bakshyRealworldVideoAdaptation2019,bartulovicBiasesDataDrivenNetworking2017,yanPantheonTrainingGround2018}.
In response, research has focused on providing
better model architectures~\cite{abbaslooClassicMeetsModern2020,jayDeepReinforcementLearning2019,yanLearningSituRandomized2020} and training algorithms~\cite{xiaGenetAutomaticCurriculum2022},
model ensembles for predictions and active learning~\cite{mallickMatchmakerDataDrift2022,heAutoMLSurveyStateoftheart2021},
real-world evaluation platforms~\cite{yanPantheonTrainingGround2018, yanLearningSituRandomized2020},
uncertainty estimation~\cite{rotmanOnlineSafetyAssurance2020}
and model verification~\cite{eliyahuVerifyingLearningaugmentedSystems2021}.

A better sample selection is beneficial to all these advances.
\name is orthogonal to and complements these works, opening an exciting potential.
We show in \cref{ssec:benchmarks} that combining \name with JTT or Matchmaker improves performance further: once \name decides to retrain, we train better with JTT, and MatchMaker benefits from an ensemble of \name-trained models.
However, optimizing training or model architectures is beyond the scope of this work, which focuses on \emph{identifying the most valuable samples} for retraining and deciding \emph{when to retrain.}
We look forward to future research investigating, \eg how to design ML models to best leverage coverage maximization.

\paragraph{Ethical issues} This work does not raise any ethical issues.

\message{^^JLASTBODYPAGE \thepage^^J}

\bibliography{2023_Memento}
\bibliographystyle{plainurl}

\message{^^JLASTREFERENCESPAGE \thepage^^J}

\ifbool{includeappendix}{%
    \clearpage
    \appendix
    \balance

\section{Artefacts and Replication.}
\label{app:artefacts}

All code to reproduce the results in this paper is publicly available and provided as an archive in the supplemental material of this paper for review.
The archive contains a README with specific instructions on how to run the code.
Below, we give an overview of reproducing the Puffer and synthetic results, and how to use Memento in other projects.

\paragraph{Puffer}
All Puffer-based experiments are based on the data published by the Puffer project~\cite{Puffer}. To replicate our results, you need to first download the Puffer data with the commands provided in the README file.
It is possible to download only a subset of the published data to save both space and computation time.
In particular, we provide a small demonstration script that uses the same code as our main experiments but uses a week of data.

Once data is downloaded, you can run two kinds of commands, both of which are described in the README file: analysis and replay.
The analysis aggregates the downloaded data to create the figures shown in \cref{ssec:deployment}, comparing the performance of \name to \fugu.
The replay retrains the TTP model with data selected by \name to create the figures shown in \cref{ssec:benchmarks}, evaluating different parameter choices with different periods of data to ensure that results are replicable.

\paragraph{Synthetic}
All synthetic experiments are based on data generated by an ns-3 network simulation. We provide the data used in the paper as well as the simulation scripts to (re-)generate it.
The README file describes the commands to replay the simulation data for retraining with \name to create the figures shown in \cref{sec:simulation}.

\paragraph{Memento}
We provide the implementation of \name as a Python package so it can be easily included in other projects.
Our implementation offers a clean interface for instantiating a memory with a given capacity, which exposes methods to \emph{insert} and \emph{get} samples.
We provide implementations for the distance metrics used in this paper, as well as a generic interface to implement new metrics.
Finally, we also implement alternative sample selection strategies with the same interface, such as random selection, loss-based selection, or LARS (see \cref{sec:simulation}), as well as an interface to combine multiple memories.

\section{Puffer: Supplemental Results}
\label{app:puffer_comparison_extra}

\paragraph{Recurring patterns}
\cref{fig:appendix_batch_coverage} illustrates that the Puffer traffic does contain patterns that recur at the tail.
Each line in this plot corresponds to one batch in the memory assembled by \name after three weeks of sample selection.
The lines show the density of batches with respect to the daily samples over the next 200 days; this captures ``how similar the batches in memory are to the traffic of the current day.''
It shows that some tail batches (the lines with the lowest densities) are sometimes more represented in the daily traffic for a couple of days (see \eg the step around week 7), then fade away.

\begin{figure}[h]
  \centering
  \includegraphics[scale=0.5]{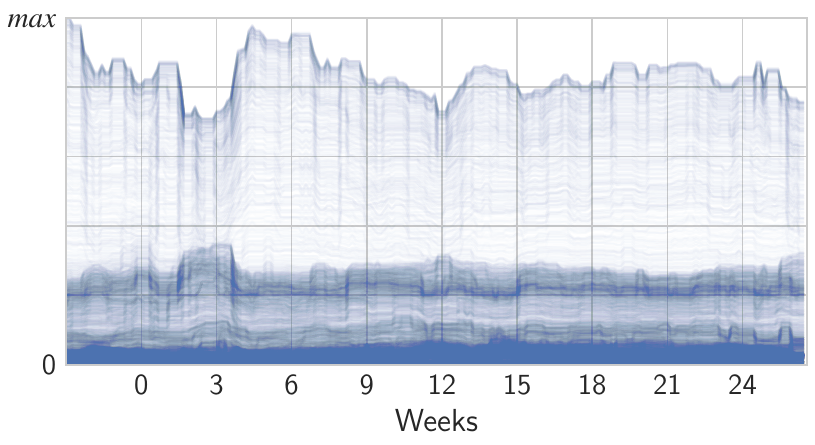}
  \caption{Coverage of batches in the daily samples}
  \label{fig:appendix_batch_coverage}
\end{figure}

\paragraph{Aggregate performance}
\cref{fig:appendix_aggregated} shows the aggregated SSIM and percent of stream-time spent stalled on a 2D grid in the style of the Puffer~\cite{yanLearningSituRandomized2020} publication and website~\cite{Puffer}.
Concretely, we show two sets of ABRs.
First \name (default and $T=0$) and \fugufeb, aggregated since the latest version of \name was deployed on September 19th, 2022 until February 13th, the cutoff for our current evaluation.
We cannot aggregate over the deployment duration \fugu as it was discontinued on October 5th.
Thus, we additionally include a second set consisting of \fugu and \fugufeb, aggregated from 2020 until \fugu was discontinued.

\paragraph{Timeseries (real-world)}
\cref{fig:appendix_qoe} shows QoE results per day over time for \name with default parameters, \name ($T=0$) and for \fugufeb since September 19th.

\cref{fig:appendix_qoe_smooth} shows the same results, but the mean and bootstrapped 90\% CI over a two-week sliding window.

\cref{fig:appendix_logscore} shows the prediction improvements over \fugufeb for \name with default parameters and with $T=0$.

\paragraph{Past degradation of \name ($T=0$) over time}
In a previous deployment, we observed \name ($T=0$) to degrade over time, as shown in \cref{fig:appendix_long_term_degradation}.
Without forgetting, it kept accumulating noise and retraining.
At first, the Puffer control loop around the TTP was able to compensate for this degradation, but as the model became too bad, it failed.
Over time, the Image quality degraded by over \SI{30}{\percent}.

However, we later discovered an issue in this deployment that prevented \name from using the deployed models' prediction, effectively disabling BBDR.
After fixing this issue, we observed that the performance of \name ($T=0$) recovered, highlighting the benefits of considering both prediction and output space.
Nevertheless, we are again beginning to see the signs of noise accumulation, in particular stronger coverage increase and frequent retraining, and are monitoring the current \name ($T=0$) deployment closely.

\begin{figure}[h]
  \centering
  \includegraphics[scale=0.5]{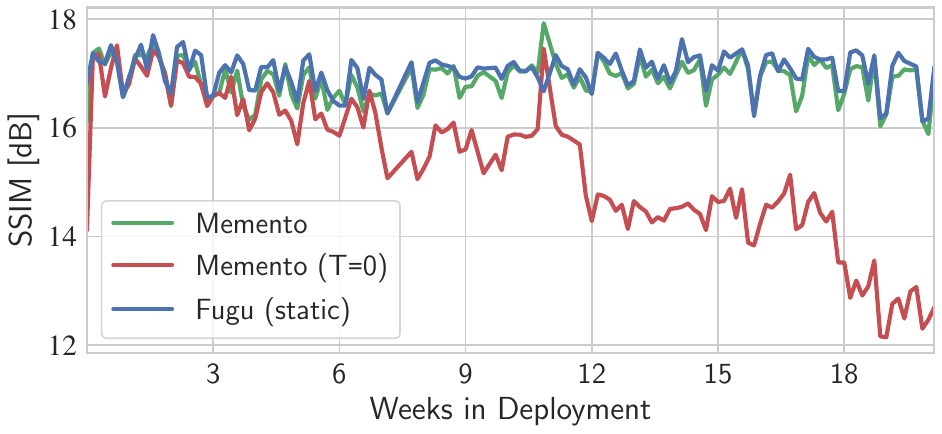}
  \caption{Long-term degradation of Memento ($T=0$) over time in a previous deployment on Puffer.)}
  \label{fig:appendix_long_term_degradation}
\end{figure}

\paragraph{Prediction score improvements compared to the past}
\cref{fig:appendix_past_scores} shows \cref{fig:score_improvement} overlayed with the prediction score improvements of \fugu compared to \fugufeb in the past.
These curves are not directly comparable, as they come from different periods of time and the underlying data may have shifted.
However, we can see that daily retraining with random samples did not consistently improve the TTP tail score; it even worsened the tail score for \SI{20}{\percent} of days. This may explain why daily retraining yields significantly smaller tail QoE improvements than retraining with samples selected by \name, which consistently improves the tail predictions.

\begin{figure}[h]
  \centering
  \includegraphics[scale=0.75]{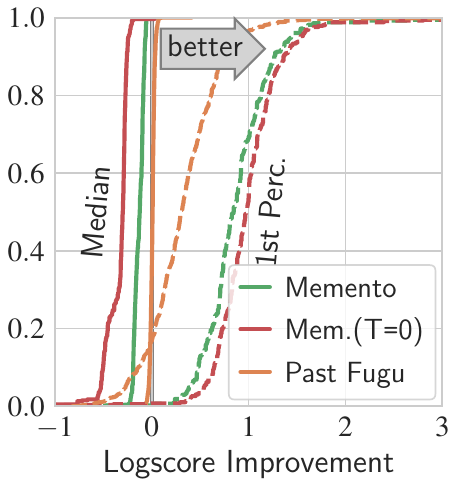}
  \caption{Retraining with daily samples did not consistently improve the tail prediction score in the past.}
  \label{fig:appendix_past_scores}
\end{figure}

\paragraph{Alternative selection metrics}
\cref{fig:appendix_alt_selection} shows further alternative sample selection metrics in addition to loss (\cref{ssec:benchmarks}).

\cref{fig:appendix_alt_confidence} shows the results for model confidence, \ie the probability of the predicted transmission time bin (Puffer separates the transmit time into 21 bins ranging from \SI{0.125}{\second} to \SI{10}{\second}). With this metric, \name prefers to discard samples with high confidence to keep `difficult' samples with low confidence.
For high temperatures, performance improvements are small.
For low temperatures, we observe strong variation between runs, with more performance degradation than improvement.
In summary, this selection metric is unreliable and fails to consistently improve performance.

\cref{fig:appendix_alt_counts} shows the results for label counts, a simplified version of density that is often used for classification data.
We use the transmission time bin of each sample as the label.
With this metric, \name prefers to discard samples with high label counts to keep `rare' samples with low label counts.
We observe this approach to drastically reduce performance.
In the Puffer environment, the majority of samples are assigned to the lowest transmit time bins. Going by label counts alone removes too many of these samples and the model forgets common patterns, similar to the loss metric.

Finally, \cref{fig:appendix_alt_stalls} shows the results for stalled sessions.
With this metric, \name prefers to discard samples if they belong to a session that did not stall to keep samples from sessions that did stall.
We observe consistent improvements, but they are small.
It does not provide a fine-grained enough selection to significantly improve tail performance.

\paragraph{Increased memory capacity}
\cref{fig:appendix_timelines_base} shows the results for \name compared to a random memory with the same capacity (both \meganum{1}) and to a random memory with a double capacity (\meganum{2}). The random memory is set up like \fugu, selecting samples randomly over the last two weeks.
We can see that simply increasing the capacity fails to address dataset imbalance, and performance is virtually identical at double the training effort.

\paragraph{Timeseries (replay)}
\cref{fig:appendix_timelines_benchmarks} shows the evolution of each benchmark experiment over the whole 6-month duration.
\cref{fig:appendix_timelines_selection} and \cref{fig:appendix_timelines_combinations} show the same time series for the experiments with alternative selection metrics, and for combining \name with Matchmaker and JTT, respectively.

For temperature-related benchmarks, we observe that a high temperature (uniformly random selection) performs slightly worse at the tail than \fugu.
This may be because \fugu keeps samples from the past \num{14} days, while a uniformly random selection phases samples out more quickly.

\pagebreak
\section{\mbox{Simulation: Supplemental Results}}
\label{app:sim_extra}

\paragraph{Simulation setup}
\cref{fig:simulation_network} illustrates the setup we used for the evaluation of \name in simulation.

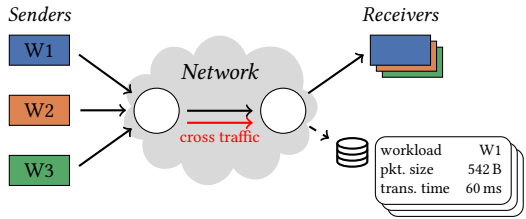
\begin{figure}[h]
  \centering
  \begin{tikzpicture}[%
        scale=0.8, transform shape,
        frontbox/.style={draw, fill=white, rectangle, minimum height=0.5cm, minimum width=1cm},
        midbox/.style={frontbox, xshift=0.1cm, yshift=-0.1cm},
        backbox/.style={frontbox, xshift=0.2cm, yshift=-0.2cm},
        router/.style={draw, circle, fill=white, minimum width=0.75cm, minimum height=0.75cm},
        arrow/.style={->, thick, shorten >=0.1cm,shorten <=0.1cm},
        sample/.style={scale=0.75, anchor=north west, align=left, rounded corners},
    ]

    \makeatletter
    \tikzset{
        database/.style={
                path picture={
                        \draw (0, 1.5*\database@segmentheight) circle [x radius=\database@radius,y radius=\database@aspectratio*\database@radius];
                        \draw (-\database@radius, 0.5*\database@segmentheight) arc [start angle=180,end angle=360,x radius=\database@radius, y radius=\database@aspectratio*\database@radius];
                        \draw (-\database@radius,-0.5*\database@segmentheight) arc [start angle=180,end angle=360,x radius=\database@radius, y radius=\database@aspectratio*\database@radius];
                        \draw (-\database@radius,1.5*\database@segmentheight) -- ++(0,-3*\database@segmentheight) arc [start angle=180,end angle=360,x radius=\database@radius, y radius=\database@aspectratio*\database@radius] -- ++(0,3*\database@segmentheight);
                    },
                minimum width=2*\database@radius + \pgflinewidth,
                minimum height=3*\database@segmentheight + 2*\database@aspectratio*\database@radius + \pgflinewidth,
            },
        database segment height/.store in=\database@segmentheight,
        database radius/.store in=\database@radius,
        database aspect ratio/.store in=\database@aspectratio,
        database segment height=0.1cm,
        database radius=0.25cm,
        database aspect ratio=0.35,
    }
    \makeatother

    \newcommand{\sample}{%
        \begin{tabular}{@{}lr@{}}
            workload    & W1                     \\
            pkt. size   & \SI{542}{B}            \\
            trans. time & \SI{60}{\milli\second} \\
        \end{tabular}
    }

    \begin{scope}
        \node {\textit{Senders}};

        \begin{scope}[yshift=-0.6cm]
            \begin{scope}
                \node[frontbox, fill=w1color] (w1s) {W1};
            \end{scope}

            \begin{scope}[yshift=-1cm]
                \node[frontbox, fill=w2color] (w2s) {W2};
            \end{scope}

            \begin{scope}[yshift=-2cm]
                \node[frontbox, fill=w3color] (w3s) {W3};
            \end{scope}

        \end{scope} %
    \end{scope} %

    \begin{scope}[xshift=3cm, yshift=-0.35cm]
        \node[
            cloud, cloud puffs=10.5, minimum width=3.25cm, minimum height=2.5cm, fill=gray!30,
            anchor=north,
        ] (network) {};
        \node at ([yshift=-0.6cm]network.north) {\large\textit{Network}};

        \node[router] at ([xshift=0.5cm]network.west) (A) {};
        \node[router] at ([xshift=-0.5cm]network.east) (B) {};
        \draw[arrow] (A) -- (B);
        \draw[arrow, red] ([yshift=-0.2cm]A.east) -- ([yshift=-0.2cm]B.west)
        node[midway, below, align=center] {\footnotesize cross traffic};
    \end{scope} %

    \begin{scope}[xshift=6cm]
        \node {\textit{Receivers}};

        \begin{scope}[yshift=-0.6cm]
            \node[backbox, fill=w3color] {};
            \node[midbox, fill=w2color] {};
            \node[frontbox, fill=w1color] (r) {};
        \end{scope}
    \end{scope} %

    \begin{scope}[xshift=6cm, yshift=-1.7cm]

        \node[database,thick] (database) at (-0.8cm, -0.6cm) {};

        \begin{scope}[shift={([xshift=0.1cm]database.north east)}]
            \node[backbox, sample] {\phantom{\sample}};
            \node[midbox, sample] {\phantom{\sample}};
            \node[frontbox, sample] {\sample};
        \end{scope}

    \end{scope}

    \draw[arrow] ([xshift=0.05cm]w1s.east) -- (A);
    \draw[arrow] ([xshift=0.05cm]w2s.east) -- (w2s.east -| A.west);
    \draw[arrow] ([xshift=0.05cm]w3s.east) -- (A);
    \draw[arrow] (B) -- (r.west);
    \draw[arrow, dashed, ->] (B) -- (database);
\end{tikzpicture}
  \caption{%
    Simulation setup.
  }
  \label{fig:simulation_network}
\end{figure}

\paragraph{Model Architecture}
We use the following parameters for the classification and regression models used in our simulation experiments.
We use supervised training and select the number of layers, neurons, and training parameters via hyperparameter optimization~\cite{headScikitoptimizeScikitoptimizeV02018}.

We have implemented both models using Keras~\cite{cholletKeras2015}.
For all layers, we use batch normalization~\cite{ioffeBatchNormalizationAccelerating2015} and ReLU activation~\cite{nairRectifiedLinearUnits2010}.
We train both networks with the Adam optimizer~\cite{kingmaAdamMethodStochastic2017} using the default decay parameters $\beta_1=0.9,\ \beta_2=0.999$.
We train for up \num{200} epochs with early stopping.

\begin{table}[h!]
  \small
  \centering
  \begin{tabular}{lrr}
    \toprule
                       & \multicolumn{2}{c}{Model}              \\
    \cmidrule(r){2-3}
    Parameter          & Classification            & Regression \\
    \midrule
    Hidden Layers      & 3                         & 4          \\
    Hidden Units       & 512                       & 362        \\
    Learning Batchsize & 512                       & 512        \\
    Learning Rate      & \num{5.91e-5}             & 0.382      \\
    \bottomrule
  \end{tabular}
  \caption{Model parameters.}
\end{table}%

\begin{figure}[h!]
  \centering
  \includegraphics[height=3.2cm]{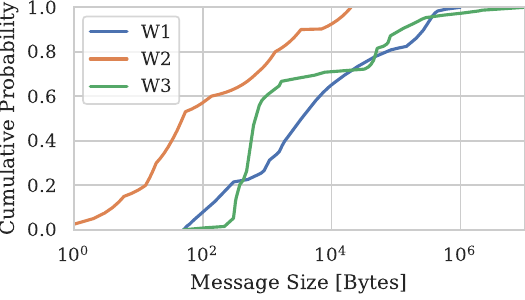}
  \caption{%
    Message size distributions~\cite{montazeriHomaReceiverdrivenLowlatency2018}.
  }
  \label{fig:simulation_distributions}
\end{figure}

\paragraph{Workload distributions}
\cref{fig:simulation_distributions} shows the message size distribution for the workloads we use, published by the HOMA project~ \cite{montazeriHomaReceiverdrivenLowlatency2018}): Facebook web server (\textbf{W1}), DC-TCP (\textbf{W2}), and Facebook Hadoop (\textbf{W3}).
Messages are generated with Poisson-distributed inter-arrival times.

\paragraph{Samples in memory}
\cref{table:sim_end_fraction} shows the number of samples in memory after the last iteration of \cref{ssec:sim_tail_perf} and \cref{ssec:sim_no_forgetting}.
\name retains all workloads in memory, whereas other approaches either forget existing or fail to pick up new samples.

\begin{table}[h!]
  \small
  \centering
  \begin{tabular}{lrrr}
\toprule
{} &    W1 &     W2 &    W3 \\
\midrule
\textit{Rare Patterns} (\cref{ssec:sim_tail_perf})\\
Memento &  5696 &   9053 &  5219 \\
Random  &   252 &  19634 &   114 \\
LARS    &  3159 &  14231 &  2610 \\
FIFO    &     0 &  20000 &     0 \\
\midrule
\textit{Incremental Learning} (\cref{ssec:sim_no_forgetting})\\
Memento &  5376 &   5888 &   8704 \\
Random  &  1266 &  17836 &    898 \\
LARS    &  9540 &   9557 &    903 \\
FIFO    &     0 &      0 &  20000 \\
\bottomrule
\end{tabular}

  \caption{%
    In-memory samples after each experiment.
  }
  \label{table:sim_end_fraction}
\end{table}

\begin{figure*}[p]
  \begin{subfigure}{0.5\textwidth}
    \centering
    \includegraphics[scale=0.75]{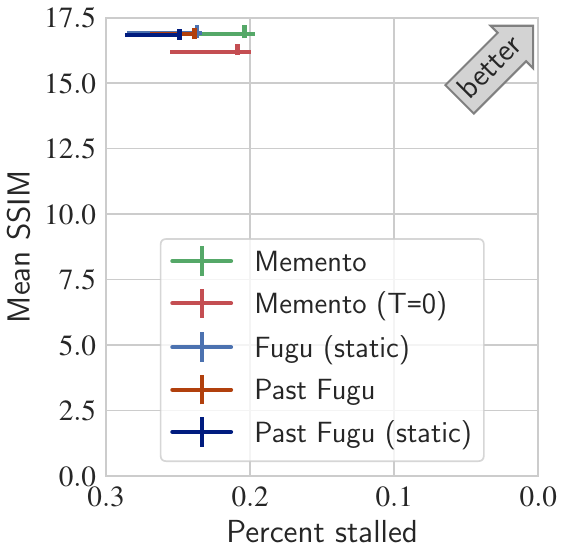}
    \caption{Image quality and fraction of stream-time spent stalled.}
  \end{subfigure}%
  \begin{subfigure}{0.5\textwidth}
    \centering
    \includegraphics[scale=0.75]{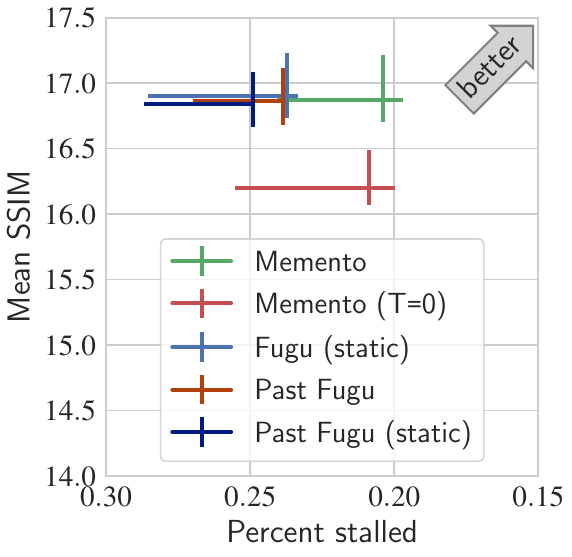}
    \caption{Zoomed in.}
  \end{subfigure}
  \caption{
    Aggregate performance. Mean and bootstrapped 90\% confidence intervals from the deployment of the current version of Memento on September 19th, 2022 until February 13th, 2023.
    ABRs annotated with `past' indicate past data from April 9th, 2020 until October 5th, 2022, when \fugu was discontinued.
  }
  \label{fig:appendix_aggregated}
\end{figure*}

\begin{figure*}[p]
  \begin{subfigure}{0.5\textwidth}
    \centering
    \includegraphics[scale=0.5]{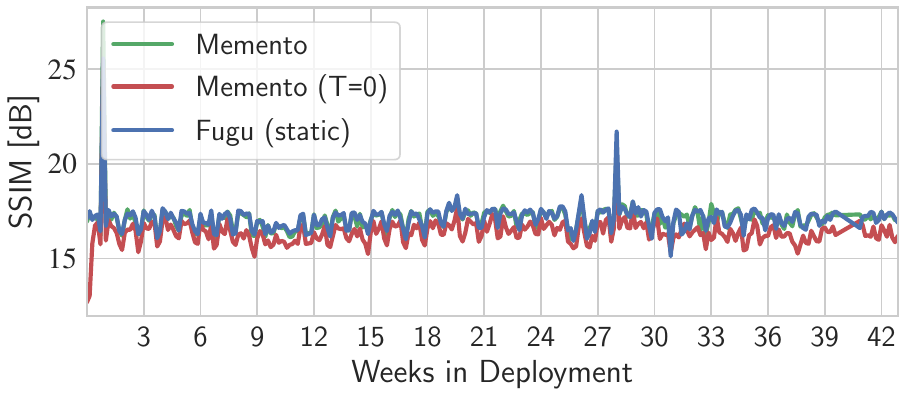}
    \caption{Image quality, measured in SSIM.}
  \end{subfigure}%
  \begin{subfigure}{0.5\textwidth}
    \centering
    \includegraphics[scale=0.5]{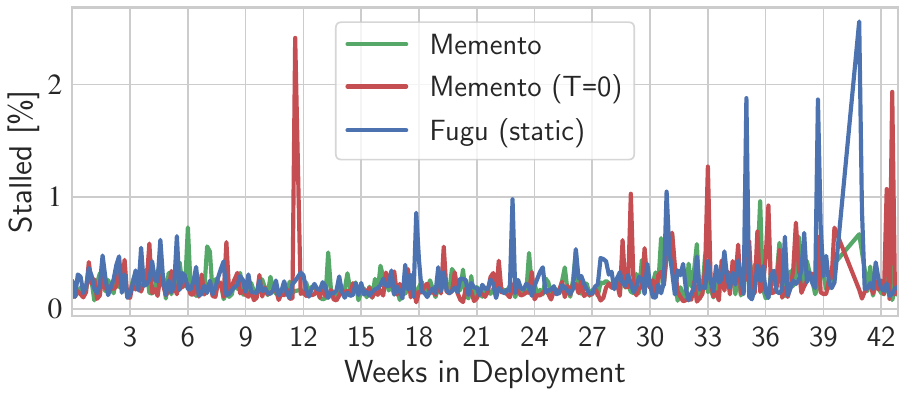}
    \caption{Percent of stream-time spent stalled.}
  \end{subfigure}
  \caption{Evolution of QoE metrics over time. Absolute Values for each ABR.}
  \label{fig:appendix_qoe}
\end{figure*}

\begin{figure*}[p]
  \begin{subfigure}{0.5\textwidth}
    \centering
    \includegraphics[scale=0.5]{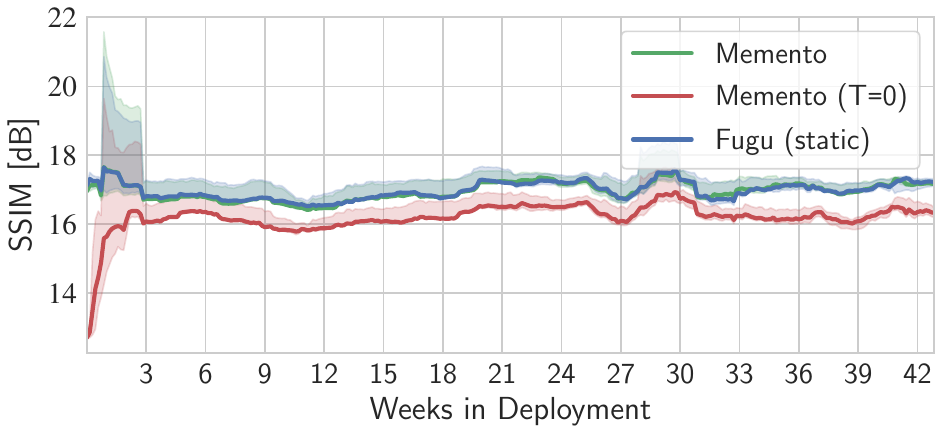}
    \caption{Image quality, measured in SSIM.}
  \end{subfigure}%
  \begin{subfigure}{0.5\textwidth}
    \centering
    \includegraphics[scale=0.5]{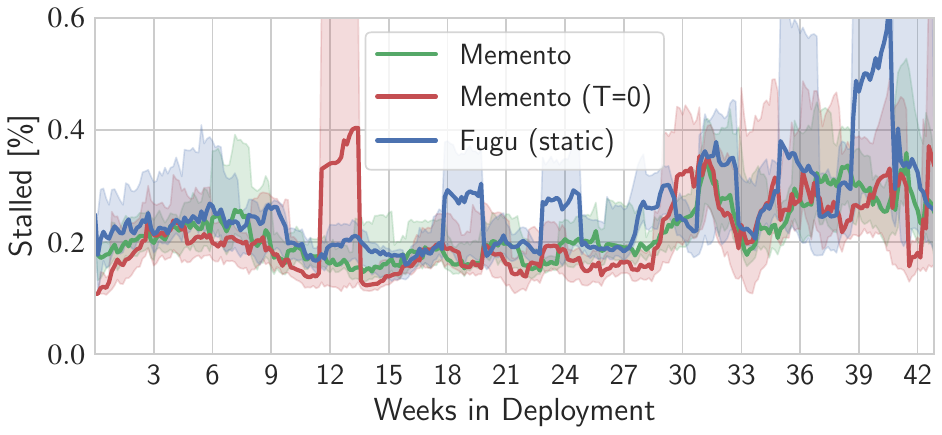}
    \caption{Percent of stream-time spent stalled.}
  \end{subfigure}
  \caption{Evolution of QoE metrics over time. Absolute Values for each ABR. Mean and bootstrapped 90\% confidence interval over two-week sliding windows.}
  \label{fig:appendix_qoe_smooth}
\end{figure*}

\begin{figure*}[p]
  \begin{subfigure}{0.5\textwidth}
    \centering
    \includegraphics[scale=0.5]{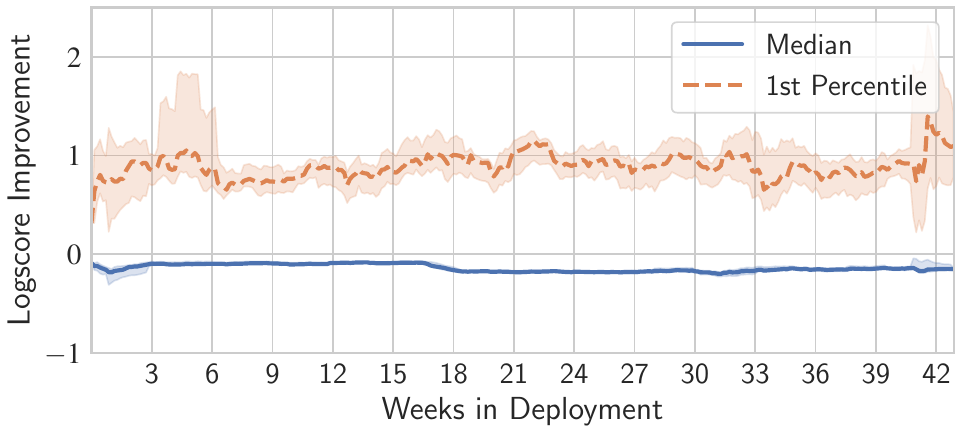}
    \caption{\name.}
  \end{subfigure}%
  \begin{subfigure}{0.5\textwidth}
    \centering
    \includegraphics[scale=0.5]{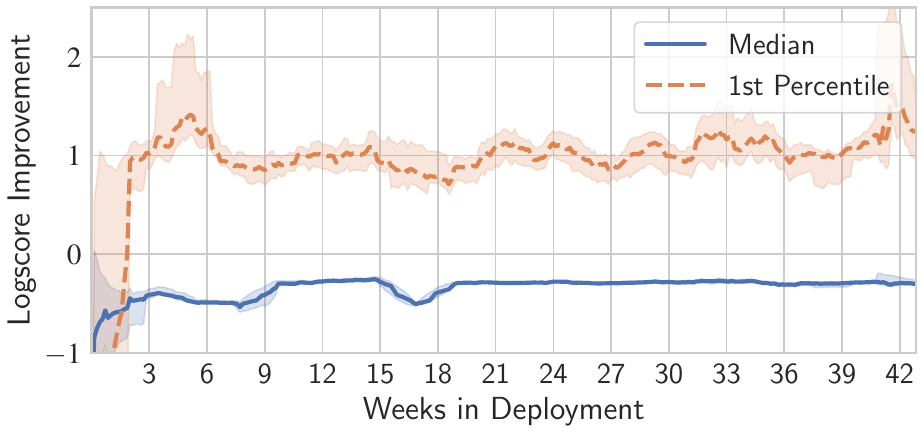}
    \caption{\name ($T=0$).}
  \end{subfigure}
  \caption{Logscore improvements compared to \fugufeb.}
  \label{fig:appendix_logscore}
\end{figure*}

\begin{figure*}[p]
  \begin{subfigure}[t]{0.32\textwidth}
    \centering
    \includegraphics[scale=0.5]{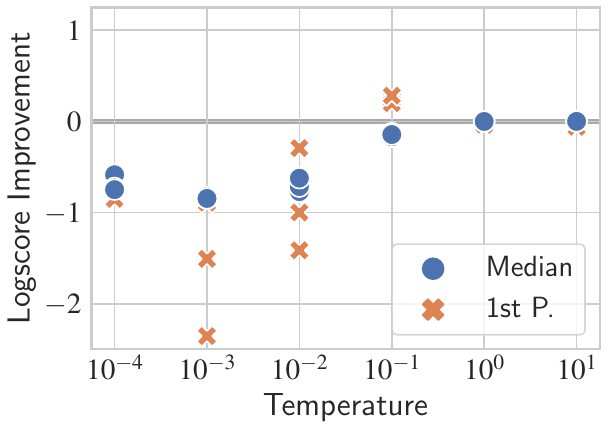}
    \caption{%
      Confidence
    }
    \label{fig:appendix_alt_confidence}
  \end{subfigure}\hfill%
  \begin{subfigure}[t]{0.32\textwidth}
    \centering
    \includegraphics[scale=0.5]{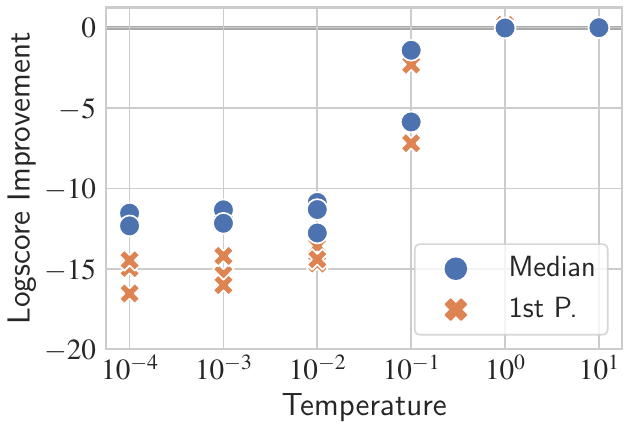}
    \caption{%
      Label counts
    }
    \label{fig:appendix_alt_counts}
  \end{subfigure}\hfill%
  \begin{subfigure}[t]{0.32\textwidth}
    \centering
    \includegraphics[scale=0.5]{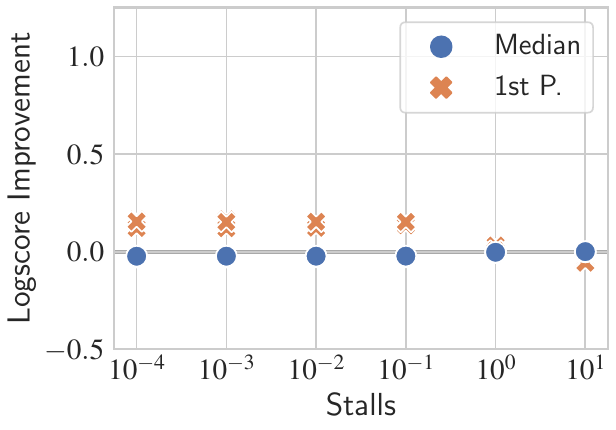}
    \caption{
      Stalled sessions
    }
    \label{fig:appendix_alt_stalls}
  \end{subfigure}
  \caption{Additional alternative selection metrics.}
  \label{fig:appendix_alt_selection}
\end{figure*}

\begin{figure*}[p]
  \begin{subfigure}{\textwidth}
    \centering
    \includegraphics[scale=0.5]{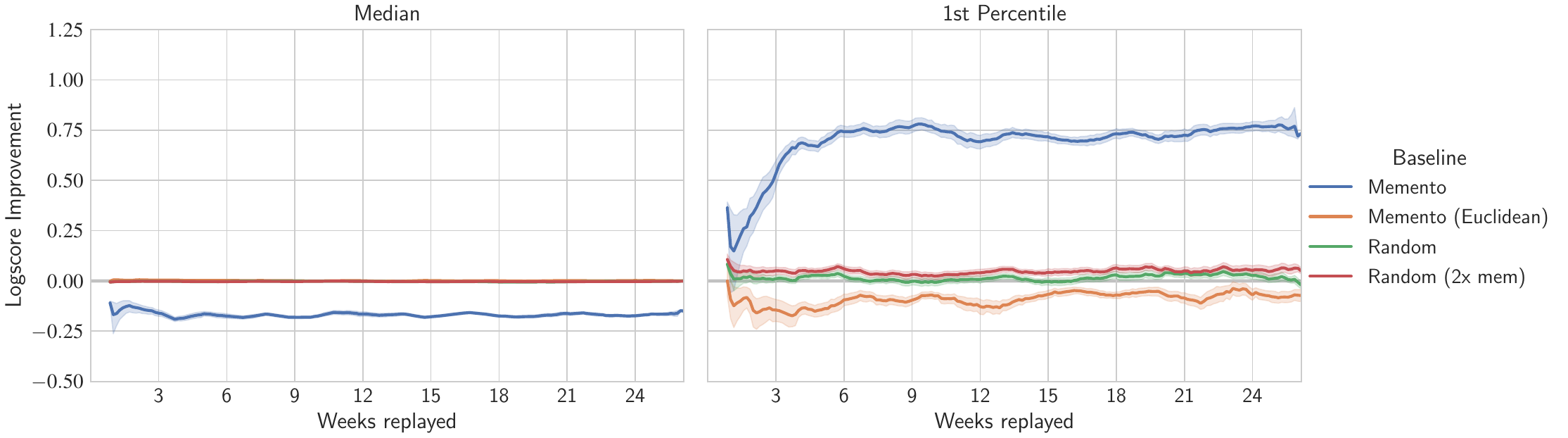}
    \caption{\name (\meganum{1} samples) compared to random sampling like \fugu (\meganum{1} and \meganum{2} samples respectively) and compared to a Memento variant using the Euclidean distance between batch averages instead of JSD.}
  \end{subfigure}\bigbreak
  \caption{Logscore improvements compared to \fugu, concretely the median and tail (1st percentile) improvements for each day. We compute the mean over all three \num{6}-month replays, and plot again the mean of this value and bootstrapped 90\% confidence intervals over two-week windows.}
  \label{fig:appendix_timelines_base}
\end{figure*}

\begin{figure*}[p]
  \begin{subfigure}{\textwidth}
    \centering
    \includegraphics[scale=0.5]{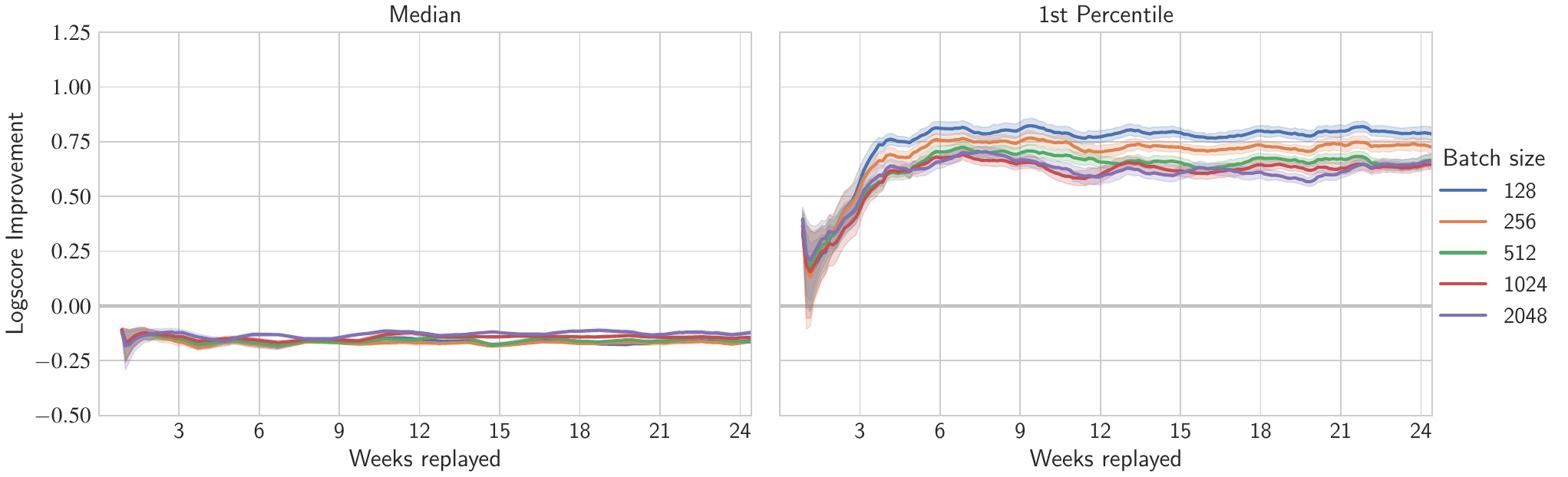}
    \caption{Batch size.}
  \end{subfigure}\bigbreak
  \begin{subfigure}{\textwidth}
    \centering
    \includegraphics[scale=0.5]{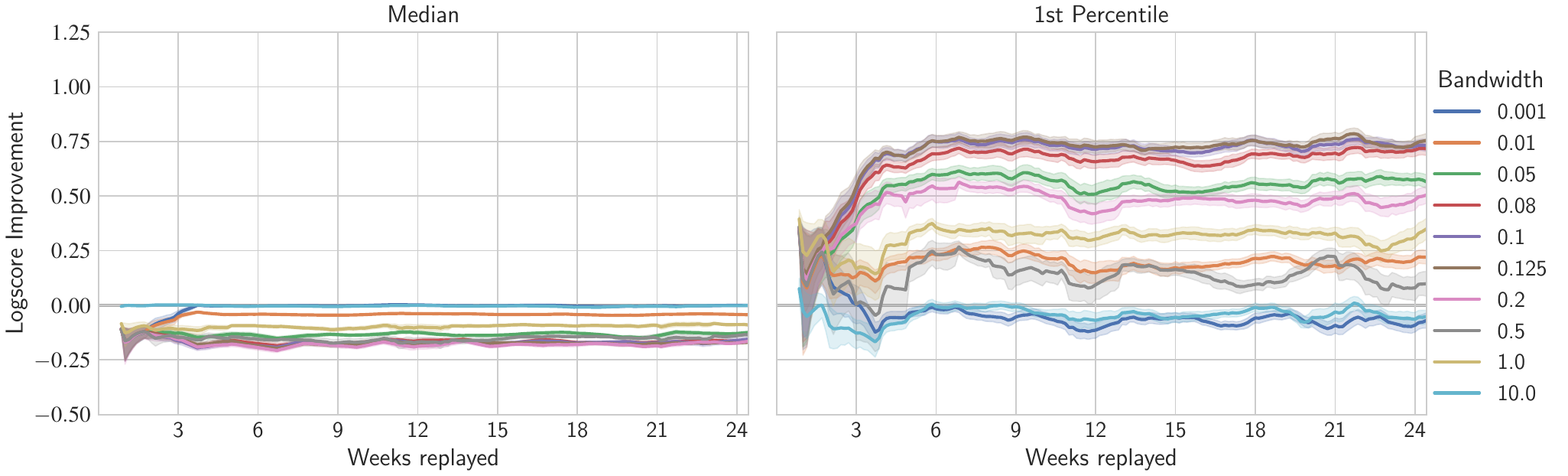}
    \caption{Bandwidth}
  \end{subfigure}\bigbreak
  \begin{subfigure}{\textwidth}
    \centering
    \includegraphics[scale=0.5]{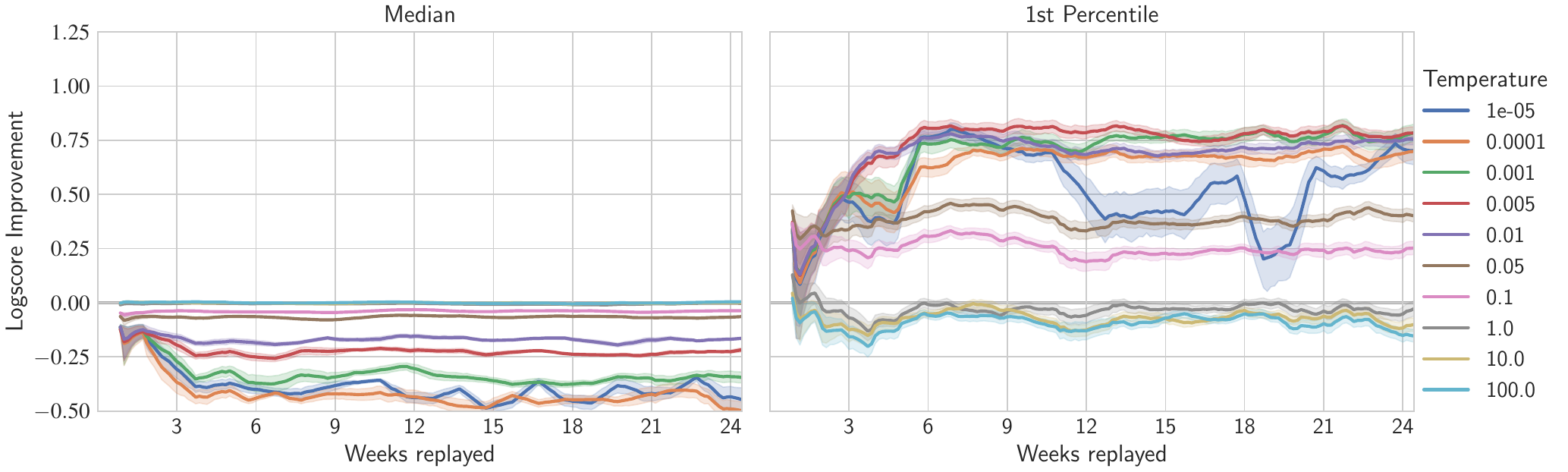}
    \caption{Temperature}
  \end{subfigure}\bigbreak
  \caption{Logscore improvements compared to \fugu (see ~\cref{fig:appendix_timelines_base}) for all selection metric experiments.}
  \label{fig:appendix_timelines_benchmarks}
\end{figure*}

\begin{figure*}[p]
  \begin{subfigure}{\textwidth}
    \centering
    \includegraphics[scale=0.42]{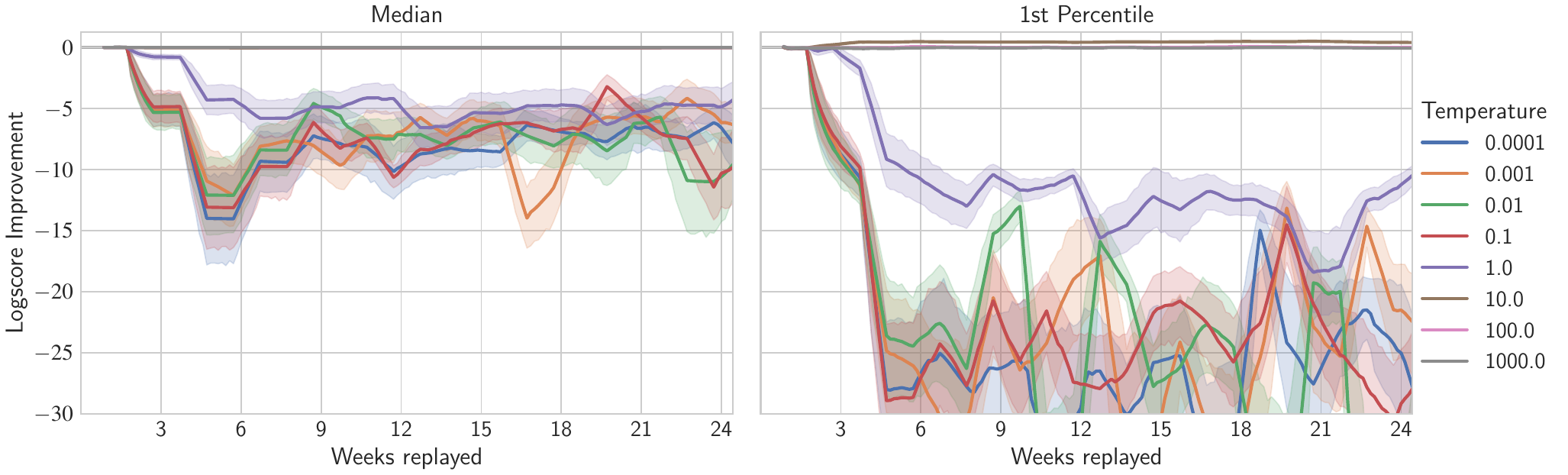}
    \caption{Loss.}
  \end{subfigure}\medbreak
  \begin{subfigure}{\textwidth}
    \centering
    \includegraphics[scale=0.42]{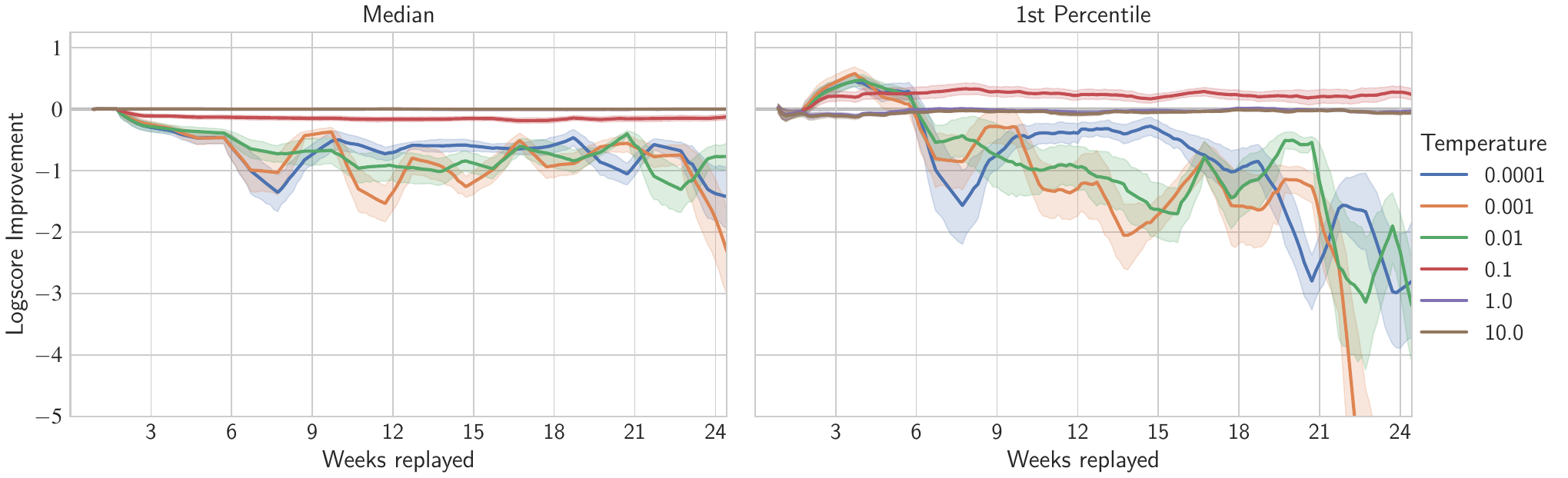}
    \caption{Confidence.}
  \end{subfigure}\bigbreak
  \begin{subfigure}{\textwidth}
    \centering
    \includegraphics[scale=0.42]{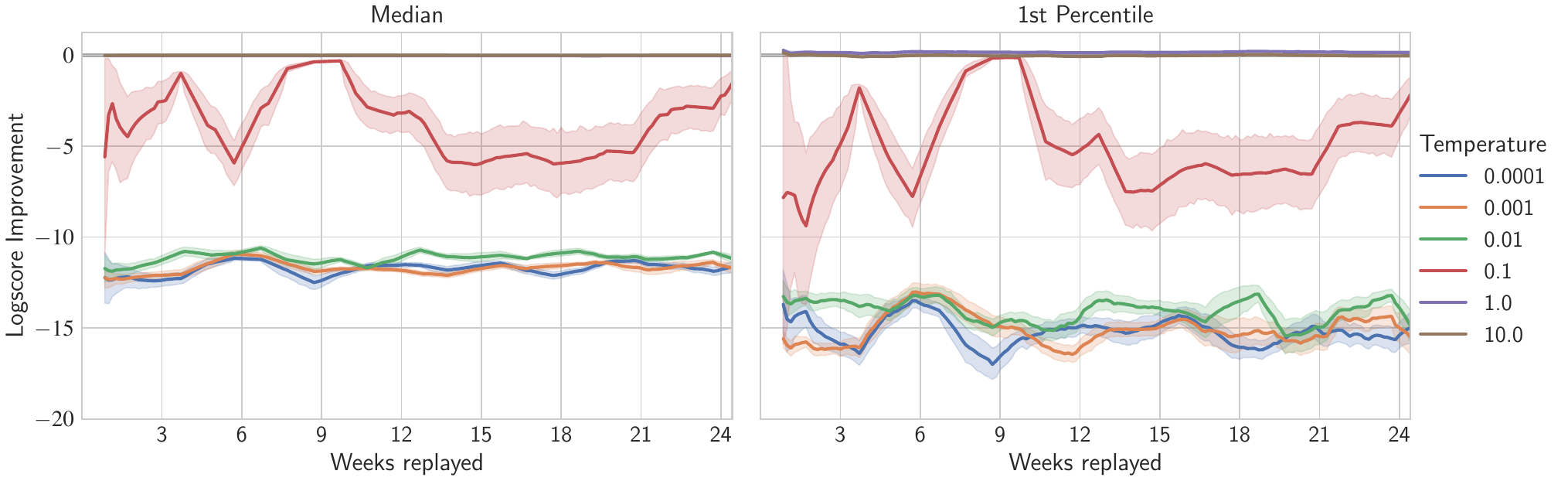}
    \caption{Label counts}
  \end{subfigure}\medbreak
  \begin{subfigure}{\textwidth}
    \centering
    \includegraphics[scale=0.42]{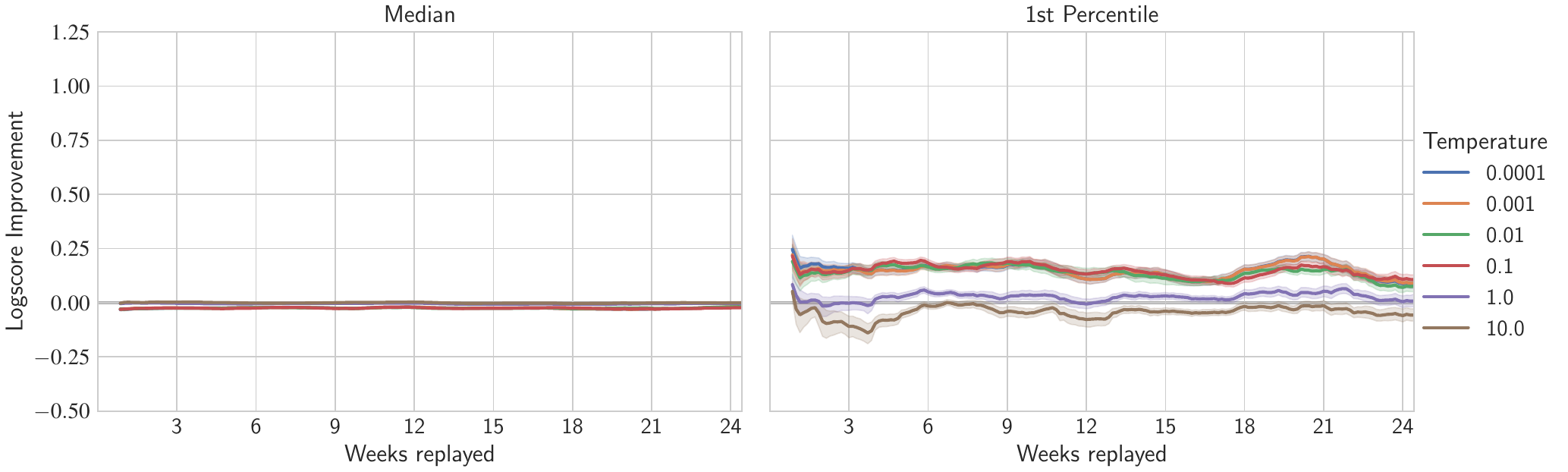}
    \caption{Stalled sessions}
  \end{subfigure}\medbreak
  \caption{Logscore improvements compared to \fugu (see ~\cref{fig:appendix_timelines_base}) for all selection metric experiments.}
  \label{fig:appendix_timelines_selection}
\end{figure*}

\begin{figure*}[p]
  \begin{subfigure}{\textwidth}
    \centering
    \includegraphics[scale=0.42]{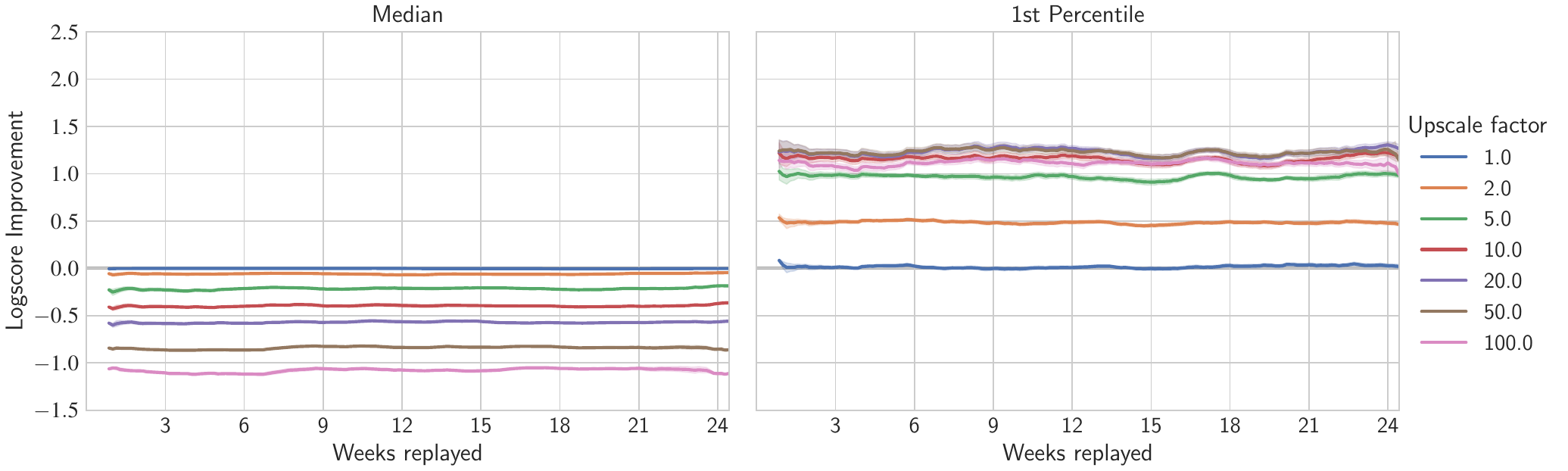}
    \caption{JTT with random sample selection.}
  \end{subfigure}\bigbreak
  \begin{subfigure}{\textwidth}
    \centering
    \includegraphics[scale=0.42]{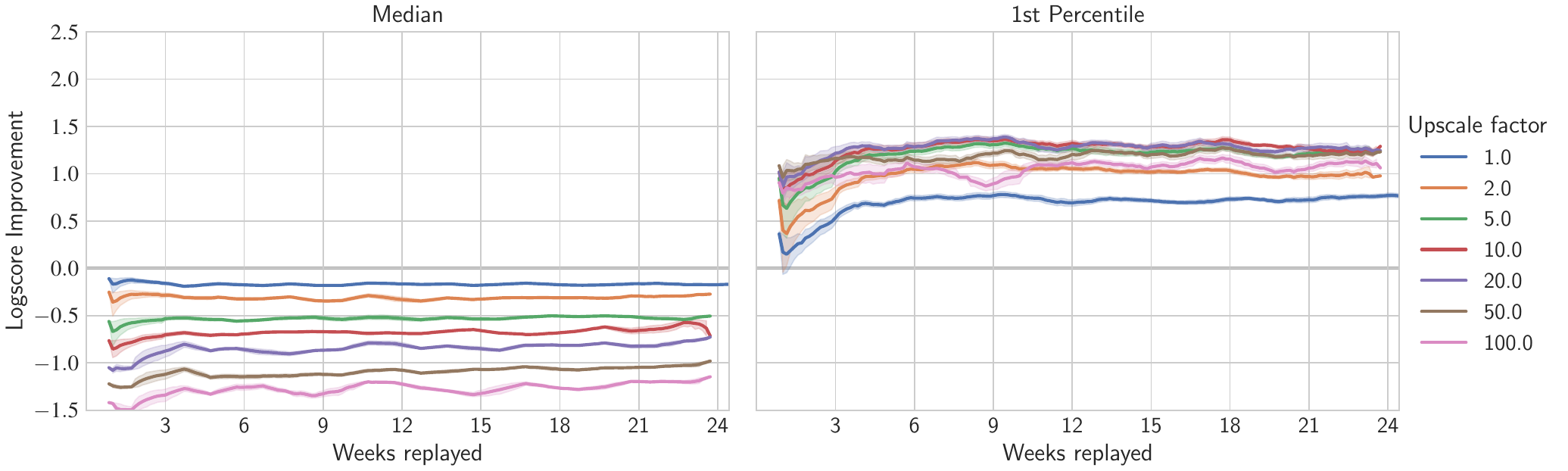}
    \caption{JTT with \name's sample selection.}
  \end{subfigure}\medbreak
  \begin{subfigure}{\textwidth}
    \centering
    \includegraphics[scale=0.42]{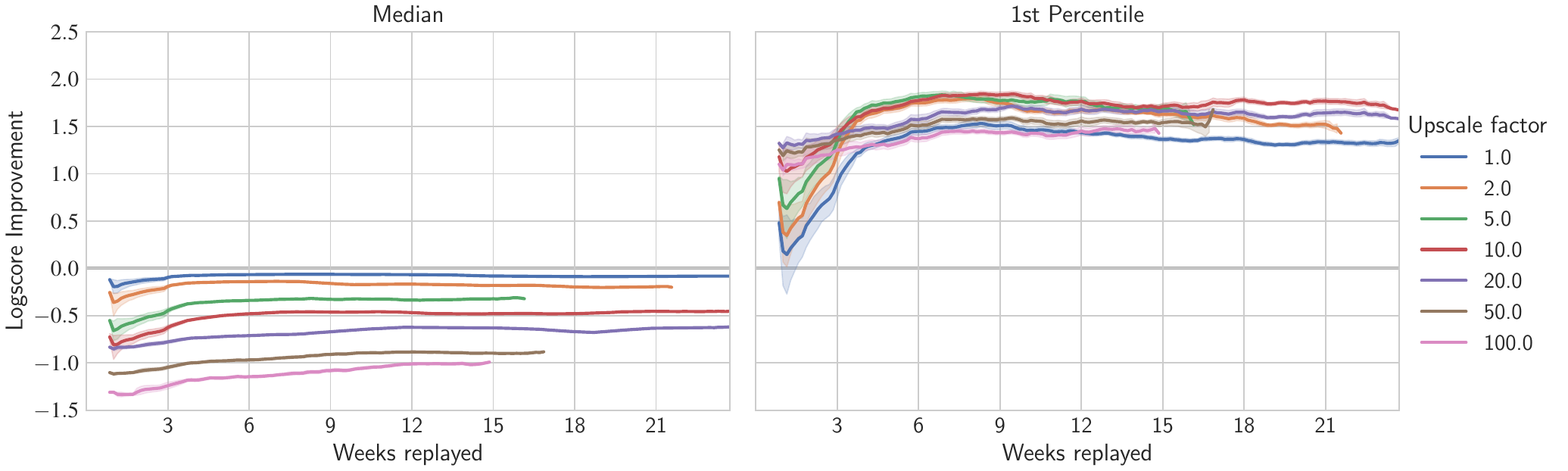}
    \caption{JTT with \name's sample selection and optimal Matchmaker predictions.}
  \end{subfigure}\medbreak
  \begin{subfigure}{\textwidth}
    \centering
    \includegraphics[scale=0.42]{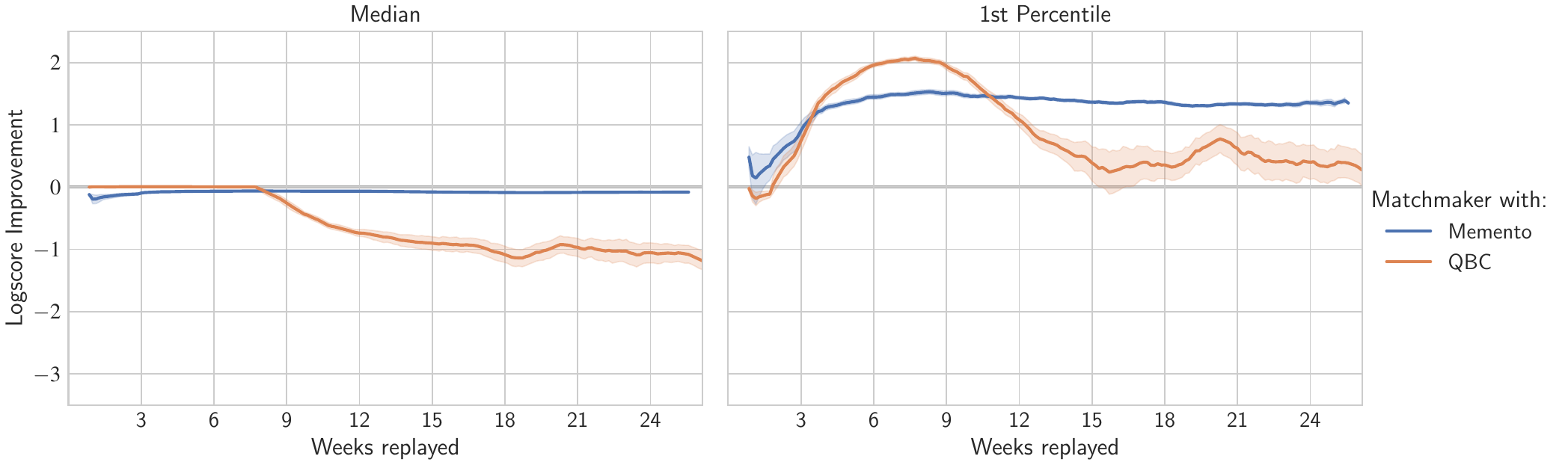}
    \caption{Memento compared to Query-By-Committee (QBC) sample selection, both with optimal Matchmaker predictions.}
  \end{subfigure}\medbreak
  \caption{Logscore improvements compared to \fugu (see ~\cref{fig:appendix_timelines_base}) for all combination experiments.}
  \label{fig:appendix_timelines_combinations}
\end{figure*}

}{}

\end{document}